%% file: paper.tex
\documentclass[12pt]{article}
\usepackage{graphicx}
\usepackage{natbib} %comment out if you do not have the package
\usepackage{url} % not crucial - just used below for the URL 
\usepackage{float}

\usepackage{hyperref}
\usepackage{url}
\usepackage{tabularx}
\usepackage{subfig}
\usepackage{xcolor}

\newcommand{\blind}{1}

\usepackage{xysong}
\usepackage{algorithm}
\usepackage{algorithmic}
\usepackage{booktabs}

\begin{document}

\def\spacingset#1{\renewcommand{\baselinestretch}%
{#1}\small\normalsize} \spacingset{1}

%%%%%%%%%%%%%%%%%%%%%%%%%%%%%%%%%%%%%%%%%%%%%%%%%%%%%%%%%%%%%%%%%%%%%%%%%%%%%%
\def\blind{0}
\if0\blind
{
  \title{\bf Quality-Ensured In-Situ Process Monitoring with Deep Canonical Correlation Analysis}\vspace{-0.1cm}
  {\author{Xiaoyang Song\hspace{.2cm}\\
    Department of Industrial and Operations Engineering, University of Michigan\\
    and\\
    Wenbo Sun \\
    University of Michigan Transportation Research Institute\\
    and\\
    Metin Kayitmazbatir\\
    Mechanical Engineering Faculty, Istanbul Technical University\\
    and\\
    Jionghua (Judy) Jin \\
    Department of Industrial and Operations Engineering, University of Michigan}
        \date{\vspace{-2ex}}
  \maketitle
} \fi

\if1\blind
{
  \bigskip
  \bigskip
  \bigskip
  \begin{center}
    {\bf \LARGE Quality-Ensured In-Situ Process Monitoring with Deep Canonical Correlation Analysis}
\end{center}
  \medskip
} \fi

% \blind

% \bigskip
% \vspace{-0.8cm}
\begin{abstract}
This paper proposes a deep learning-based approach for in-situ process monitoring that captures \textbf{nonlinear} relationships between in-control \textbf{high-dimensional} process signature signals and offline product quality data. Specifically, we introduce a Deep Canonical Correlation Analysis (DCCA)-based framework that enables the joint feature extraction and correlation analysis of \tb{multi-modal} data sources, such as optical emission spectra and CT scan images, which are collected in advanced manufacturing processes. This unified framework facilitates online quality monitoring by learning quality-oriented representations without requiring labeled defective samples and avoids the non-normality issues that often degrade traditional control chart-based monitoring techniques. We provide theoretical guarantees for the method’s stability and convergence and validate its effectiveness and practical applicability through simulation experiments and a real-world case study on Direct Metal Deposition (DMD) additive manufacturing.
\end{abstract}

\noindent%
{\it Keywords:} In-situ process monitoring, Multi-modal feature extraction, Deep canonical correlation analysis.
\vfill

\newpage
\spacingset{2} % DON'T change the spacing!
%
% \blind
\input{src/intro}

\section{Methodology}\label{sec: method}

\input{src/method}

\input{src/theory}

\input{src/simulation}

\input{src/case}

\input{src/conclusion}

\bibliography{reference}
\bibliographystyle{apalike}

\input{src/appendix}

\end{document}

%% file: src/intro.tex
\section{Introduction}\label{sec: intro}
In advanced manufacturing, in-situ process monitoring has been increasingly demanded to make timely process correction decisions to reduce unnecessary waste due to delayed inspections. While online machine vision systems are increasingly adopted for in-situ inspection of products' dimensions and surface defects, there are still challenges in in-situ assessing other critical quality characteristics during production. For instance, assessing the tensile strength of joints during welding like ultrasonic welding or friction stir blind riveting (FSBR) \citep{zerehsaz2019tool, gao2022sensor} is infeasible, as those typically require destructive tensile tests conducted offline. Similarly, detecting inner porosity defects during printing in additive manufacturing (AM) often requires expensive CT scans \citep{montazeri2020process, sun2022situ}, which are neither cost-effective nor scalable for real-world production.

The adoption of advanced sensors in manufacturing enables the real-time collection of process operating or response signals, which are often referred to as process signature signals. Those process signature signals can be utilized for process monitoring, in-situ product quality prediction, and defect detection. For instance, real time power signals and acoustic signals in ultrasonic welding \citep{guo2016online}, penetration force and thermal images during FSBR operation \citep{zerehsaz2019tool, gao2022sensor} are highly sensitive to joint quality, rendering it possible to use those process signature signals to predict the tensile strength of joints during production. Optical emission spectroscopy generated during material-laser interactions can be analyzed to detect inner porosity defects in AM processes \citep{mazumder2015design, montazeri2020process, sun2022situ}. These online process sensing signals offer a feasible and cost-effective approach for in-situ process monitoring and quality control. 

Developing quality-oriented process monitoring methods involves collecting training data under normal production conditions, including both process signature signals and their corresponding quality inspection data.
\begin{figure}[t]
    \centering
    \includegraphics[width=1.0\linewidth]{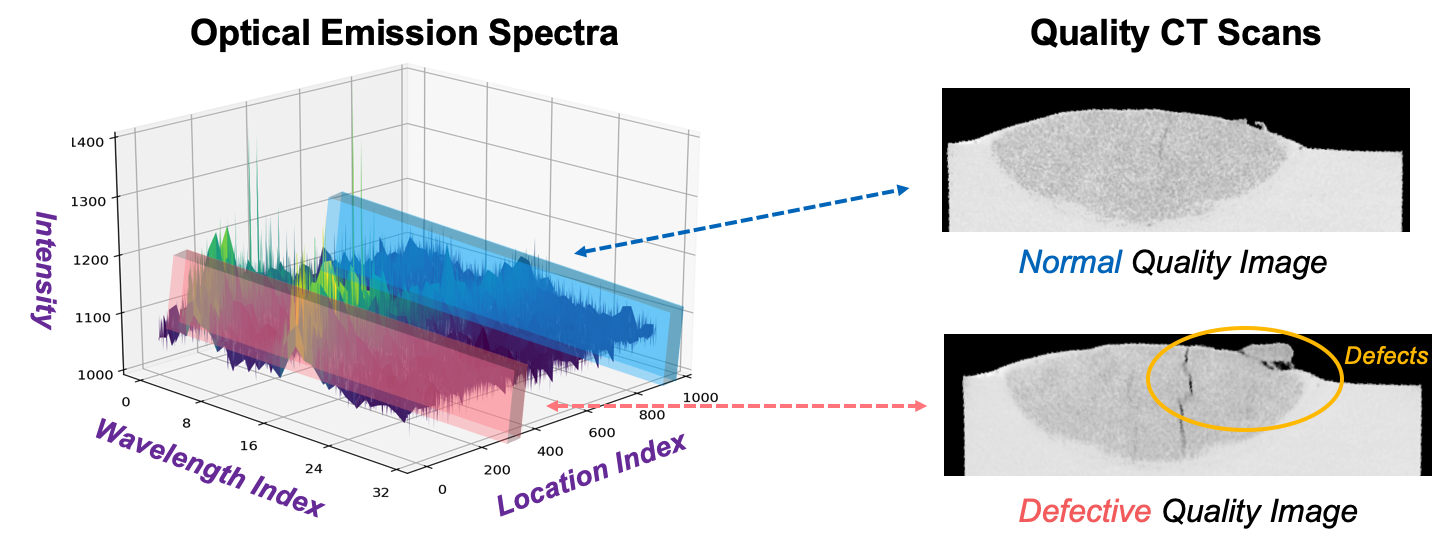}
    \caption{Examples of normal and defective spectra and CT scan pairs from a real-world metal AM process, where each CT scan image serves as the quality metric for its corresponding spectral process signal at the same printing location. More details are presented in Section \ref{sec: case-study}.}
    \label{fig: example}
\end{figure}
Given that product quality data generally exhibits a complex nonlinear dependence on process signature signals, this paper aims to develop an effective algorithm to uncover this nonlinear correlation as monitoring statistics, enabling in-situ monitoring to distinguish normal and defective parts based solely on process signals. Figure \ref{fig: example} (left) provides an example for online optical emission spectra signals collected during a metal AM operation, as well as their corresponding quality inspection images collected from offline CT scans. In general, the number and size of produced porosity sensitively affect the strength and durability of the product. Instead of relying on binary labels or discretized quality categories that are tied with specific applications \citep{montazeri2020process}, we employ offline CT scan images as generic continuous quality metrics. This approach maximally preserves product quality-related information and improves generalizability across different products' requirements, as categorical definitions and labeling criteria often vary in different applications. In concept, the dependent relationships between spectra signals and products' CT scans are different under normal and fault process operations.

Furthermore, online process signature signals, such as spectra, are high-dimensional and exhibit non-stationary spatiotemporal correlations within each operation cycle at different printing locations, which necessitates a tensor representation. Similarly, quality inspection CT images are represented in tensors, rendering it challenging to uncover nonlinear dependencies between the two complex tensor datasets. As demonstrated in our case study (Section \ref{sec: case-study}), a stream of process signature spectra signals is typically aligned with a stream of CT scan images, further highlighting the importance of tensor representations. These render feature extraction essential to capture the strong spatiotemporal correlations in process signals as well as the sparsity structures in quality images.

There are tremendous works in the literature on feature extraction of process sensing signals for process change detection \citep{colosimo2024statistical}. For instance, in the unsupervised approaches, PCA-based methods are often combined with regularization techniques such as LASSO \citep{tibshirani1996regression} to identify important features that significantly contributed to process changes for monitoring \citep{grasso2017process, colosimo2018spatially, yan2018real, ebrahimi2018large, zhang2020correlation}. However, since the basic principle of these methods is to capture and detect abnormal process variations in the data, they often suffer from the issue that the extracted features may not accurately represent the actual product defects due to the lack of supervision.

Besides the aforementioned unsupervised approaches, supervised feature extraction methods have been explored to effectively identify key features of sensing signals. There are two common strategies in this context. The first approach is model-based, which directly builds a prediction model of quality responses based on process sensing signals. This can involve constructing regression models on Key Product Characteristics (KPCs) using low-dimensional features of process signature signals as predictors \citep{zerehsaz2019tool, yan2022real, colosimo2024statistical} or developing a classifier for defect detection by leveraging modern machine learning techniques \citep{bugatti2022towards, gonzalez2020convolutional, li2020quality, bui2018monitoring, sun2022situ, guo2019profile, guo2016online, shao2013feature}. However, the extracted features are often highly sensitive to the choice of regression or classifier model structure as well as the definition of quality specifications. This renders a poor generalizability of those extracted features for different product requirements, which is a common disadvantage in these supervised approaches.
\begin{figure}
    \centering
    \includegraphics[width=1\linewidth]{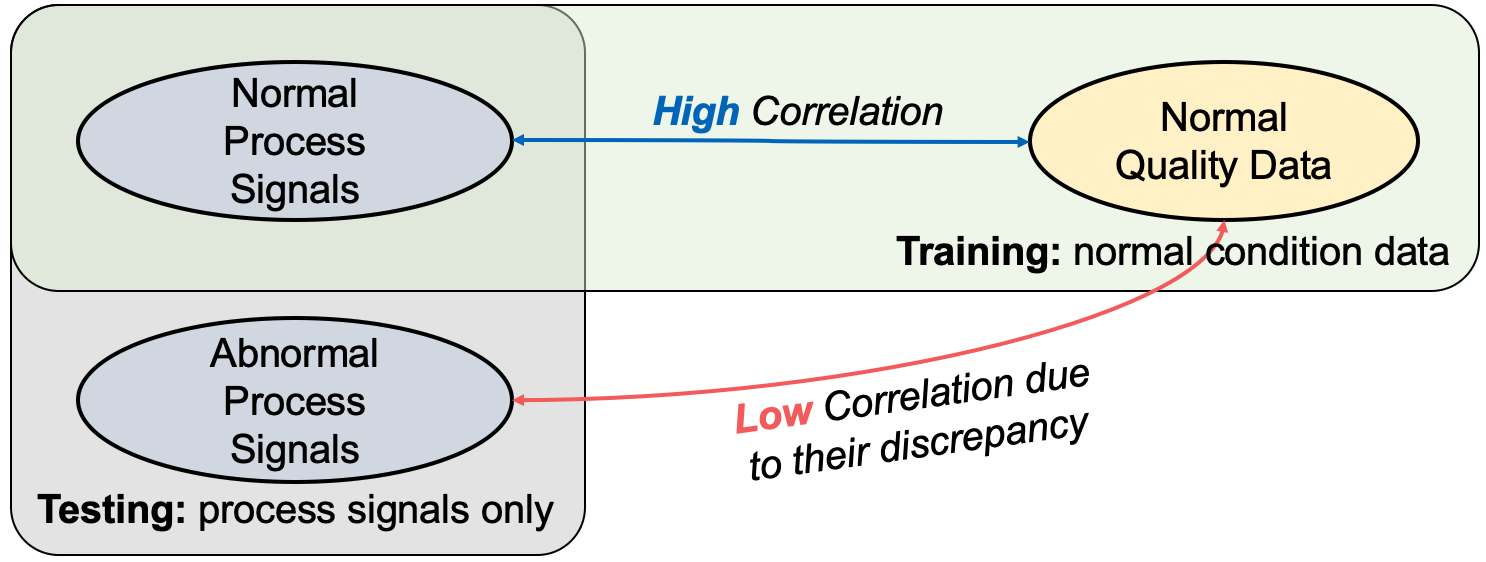}
    \caption{An overview of the proposed method and intuition.}
    \label{fig: high-level-intro}
    \vspace{-0.10in}
\end{figure}

In contrast to the model-based supervised feature extraction method, the model-free approach extracts features by maximizing the dependency between online process sensing signals and product KPCs. For instance, Partial Least Squares (PLS) are utilized to find projection directions that maximize the covariance between process sensing signals and quality data \citep{mardia1979multivariate, song2002partial}. However, this method is less applicable because covariance estimation is typically unreliable when both process sensing signals and quality data are extremely high-dimensional and under limited sample sizes. To mitigate these issues, the Canonical Correlation Analysis (CCA) method \citep{mardia1979multivariate} seeks to maximize the linear correlation between extracted features of two datasets. To handle nonlinear dependence, improved methods like Kernel CCA (KCCA) \citep{shawe2004kernel} and Deep CCA (DCCA) \citep{andrew2013deep} have been developed, where they employed either a kernel function or a deep neural network (DNN) to learn nonlinear data representations, respectively. Leveraging the flexibility of deep neural networks, DCCA extends classical Canonical Correlation Analysis (CCA) to discover nonlinear relationships, making it particularly well-suited for feature extraction in a challenging setting involving nonlinear dependencies between two high-dimensional datasets. In this paper, we extend DCCA with an online monitoring framework for effective feature extraction and in-situ monitoring.

\begin{figure}[t]
    \centering
    \includegraphics[width=1.0\linewidth]{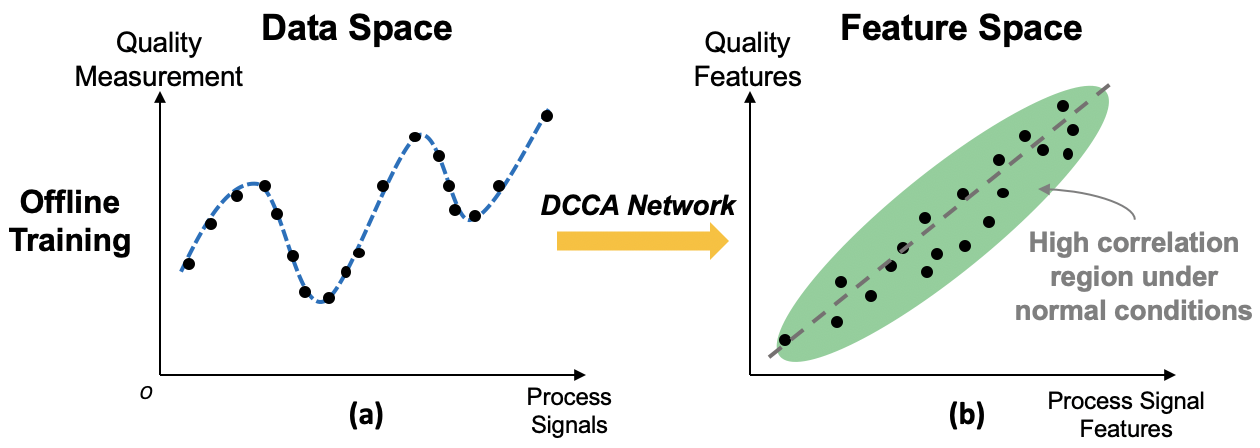}
    \caption{1D-Illustration of the objectives of the proposed offline training for the DCCA-based feature extraction model.\vspace{-0.15in}}
    \label{fig: illustration}
\end{figure}
In short, the proposed method aims to learn feature representations of both process sensing signals and quality data using DCCA with the goal of maximizing the correlation between the extracted features, in which the quality data automatically guides the training of the feature extractors. Figure \ref{fig: high-level-intro} provides a high-level overview of the proposed method. In particular, the feature extractors are trained with only data collected under the normal process conditions, which returns a high correlation score between the extracted features only when both process signature signals and quality measurement data are generated from the normal conditions, as shown in Figure \ref{fig: illustration} (a) and (b). For online monitoring during the testing stage, rather than constructing a traditional $T^2$ control chart based on the extracted features, we utilize correlation models to further project the features into the correlation score space, enabling effective separation of in-control and out-of-control conditions, even though the extracted features are not normally distributed. Specifically, when a new spectra signal is observed, it will be paired with the closest reference CT scan of the normal condition chosen from the training set, with which their correlation score will be evaluated to perform process monitoring. This additional augmentation and projection step effectively avoids the common non-normality issues often encountered when applying $T^2$ chart to real-world data, which typically cause significant performance degradation.

In summary, this paper proposes a model-free feature extraction approach for high-dimensional tensorial in-situ sensing signals and offline quality data, aiming to uncover their nonlinear dependencies. Specifically, the proposed method extracts strongly correlated features and combines them with a heuristic searching algorithm for efficient in-situ process monitoring, which avoids the normality assumption on the extracted features. The remainder of the paper is organized as follows. In Section \ref{sec: method}, we present the entire framework in detail, including mathematical formulation and the detailed training procedures. We then provide a rigorous theoretical guarantee to justify the effectiveness of the proposed method in an empirical setting in Section \ref{sec: theory}. A simulation study in Section \ref{sec: simulation} and a case study in Section \ref{sec: case-study} using a real-world manufacturing dataset are conducted to demonstrate its superior performance over other traditional $T^2$-chart-based monitoring methods and its practical applicability.

%% file: src/method.tex
\subsection{Problem Formulation}\label{ssec: problem-formulation}
In general, the goal of in-situ process monitoring is to assess the quality of the production process using only observable online process signature signals. Let $(\bx, \by) \in \reals^{d_1 + d_2}$ denote the pair of random variables of online process signature signals and offline quality measurements that are unobservable during production, with $d_1$ and $d_2$ representing their vectorized data dimensions transformed from matrix or tensor forms, respectively. We further assume that all samples of $(\bx, \by)$ are independent and identically distributed, yet between $\bx$ and $\by$ they exhibit inherent cross-dependencies reflecting the characteristics of process physics. Correspondingly, let $(\bu, \bv)=(\bff(\bx), \bg(\by)) \in \reals^{2p}$ denote the transformed features for $(\bx, \by)$ using the feature extraction networks $\bff(\cdot)$ and $\bg(\cdot)$, respectively, with $p$ representing the common feature dimension. Depending on process conditions, we use $\cP_{0}$ and $\cP_{1}$ to represent the joint distributions of $(\bx, \by)$ collected from normal and abnormal process conditions, whereas $\cP_{0,x}, \cP_{0,y}$, and $\cP_{1,x}$ denote the marginal distributions for normal process signature signals, normal quality data, and abnormal process signature signals, respectively. 

To perform in-situ process monitoring, this paper aims to uncover the non-linear dependencies between online process sensing signals and offline quality data. Specifically, we strive to identify two feature extractors $\bff(\cdot)$ and $\bg(\cdot)$ for them, respectively, along with a score function $s(\cdot, \cdot)$ that evaluates the dependency of two random variables. As described in Figure \ref{fig: high-level-intro}, those extracted features are expected to possess high scores only when both process sensing signals and quality data come from historical normal process conditions. Mathematically, this can be expressed as follows:
\begin{equation}
    \max_{\bff, \bg, s} \ \bbE_{(\bx, \by) \sim \cP_0}[s(\bff(\bx), \bg(\by))]
    % , \text{ and } \min_{\bff, \bg, s} \ \bbE_{(\bx, \by) \not\sim \cP_0}[s(\bff(\bx), \bg(\by))].
\end{equation}
% \textcolor{red}{Should we use $\sim\cP_1$ in the minimization problem?}
% short answer is no: as P1 is one way to achieve minimization problem; however, in general, the condition is "not belong to P0".
In this paper, we adopt the canonical correlation analysis score \citep{mardia1979multivariate} as the score function to identify $\bff(\cdot)$ and $\bg(\cdot)$ in the offline training stage. The well-trained feature extractors can then be utilized for efficient in-situ process monitoring by combining with a nearest neighbor search heuristic, which will be introduced in Section \ref{ssec: online-monitoring}.

\subsection{Deep Canonical Correlation Analysis}
First, we recall the Canonical Correlation Analysis (CCA) formulation. Let $\bx \in \reals^{d_1}$ and $\by \in \reals^{d_2}$ denote two random vectors of dimension $d_1$ and $d_2$, respectively. The objective of CCA is to find two linear projections $\bW_1^* \in \reals^{d_1 \times p}$ and $\bW_2^* \in \reals^{d_2 \times p}$ such that the sum of correlations between each dimension of the two projected views is maximized, that is,
\begin{align*}
    \parent{\bW_1^*, \bW_2^*} \ = & \ \argmax_{\bW_1, \bW_2} \ \corr\parent{\bW_1^\top\bx, \bW_2^\top\by} = \argmax_{\bW_1, \bW_2} \ \tr\parent{\bW_1^\top\bSigma_{12}\bW_2}\\
    & s.t. \ \ \bW_1^\top\bSigma_{11}\bW_1 = \bW_2^\top\bSigma_{22}\bW_2 = \bI
\end{align*}
where $p \leq \min(d_1, d_2)$ denotes the number of projections which is also known as the CCA dimension, and $\bSigma_{12}$, $\bSigma_{11}$, and $\bSigma_{22}$ represent the ground truth cross-covariance matrix of $\bx$ and $\by$, and the covariance matrix of $\bx$ and $\by$, respectively. Here we employ $\corr(\ba, \bb)$ to denote the sum of linear correlations between each corresponding dimension of random vectors $\ba$ and $\bb$. Furthermore, according to \citet{mardia1979multivariate}, we define $\bT := \bSigma_{11}^{-1/2} \bSigma_{12} \bSigma_{22}^{-1/2}$ and then the closed-form optimal objective is given by $\norm{\bT}_{*}$, where $\norm{\cdot}_*$ denotes the matrix trace norm \citep{fan1951maximum}. Note that it is common to consider the top-$k$ projections out of the total $p$ CCA dimension \citep{andrew2013deep, wang2015deep}. In this work, we stick to the case where all projections are involved ($k=p$) to reduce the number of controllable parameters.

Considering that the linear projections $\bW_1^*$ and $\bW_2^*$ may have limited expressive power for complicated inputs such as images and high-dimensional tensors, \citet{andrew2013deep} introduces Deep CCA (DCCA), where the linear projections are replaced by well-designed neural networks to enable nonlinear transformations of $\bx$ and $\by$. Differing from the classic Canonical Correlation Analysis (CCA), DCCA focuses more on feature extractions and attempts to find two non-linear feature extractors $\bff(\cdot)$ and $\bg(\cdot)$ so that the canonical correlation coefficients of the extracted features are maximized \citep{mardia1979multivariate}:
\vspace{-0.05in}
\begin{align}
   \hspace{-0.10in}\parent{\btheta_1^*, \btheta_2^*} &= \argmax_{\btheta_1, \btheta_2} \max_{\bW_1, \bW_2} \corr\parent{\bW_1^\top \bff(\bx; \btheta_1), \bW_2^\top \bg(\by; \btheta_2)},\hspace{-0.08in}
   \label{eq:DCCA}\vspace{-0.05in}
\end{align}
where $\bff(\cdot)$ and $\bg(\cdot)$ denote the neural-network-based feature extractors that transform the input data $\bx \in \reals^{d_1}$ and $\by \in \reals^{d_2}$ to $p$-dimensional features $\bff(\bx; \btheta_1)$ and $\bg(\by; \btheta_2)$, respectively, and $\btheta_1$ and $\btheta_2$ are the two vectors that collect all trainable parameters for the neural networks. Furthermore, we set the dimensions of \(\bW_1\) and \(\bW_2\) to \(p \times p\) to ensure that the DCCA dimension matches the number of DCCA projections, thereby reducing the number of tuning parameters.

In practice, calculation of canonical correlation $\corr(\cdot,\cdot)$ in Eq.\eqref{eq:DCCA} requires estimating the covariance and cross-covariance matrix of $\bff(\bx; \btheta_1)$ and $\bg(\by; \btheta_2)$. Let $\hat\bSigma_{12}$, $\hat\bSigma_{11}$, and $\hat\bSigma_{22}$ represent the estimated cross-covariance matrix between the random vectors $\bff(\bx; \btheta_1)$ and $\bg(\by; \btheta_2)$, the covariance matrix of $\bff(\bx; \btheta_1)$ and the covariance matrix of $\bg(\by; \btheta_2)$, respectively. Let $\bU, \bV \in \reals^{p \times n}$ collect the samples of $\bff(\bx; \btheta_1)$ and $\bg(\by; \btheta_2)$, then the $n$-sample estimator of $\corr(\cdot,\cdot)$ is formally given as a function $H_{n,p}: \reals^{2p \times n} \mapsto \reals$, which is defined as:
% \vspace{-0.03in}
\begin{equation}
    H_{n,p}(\bU, \bV) = \norm{\hat\bSigma_{11}^{-1/2} \hat\bSigma_{12} \hat\bSigma_{22}^{-1/2}}_*,
    \vspace{-0.05in}
\end{equation}
where $\norm{\cdot}_*$ denotes the matrix trace norm.

\subsection{DCCA-based Offline Model Training}\label{ssec: offline-training}
The first step involves training a DCCA model \citep{andrew2013deep} on a baseline dataset collected from normal process conditions such that it provides high canonical correlations when pairs of normal process sensing signals and normal quality data are presented to the model.
Let $\cX_0, \cX_1, \cY_0$, and $\cY_1$ denote the normal and abnormal observations for process sensing signals and quality data, respectively. As mentioned in the preceding section, it is typical to consider a set of observations for a decent estimation of canonical correlation coefficients during the optimization process. In particular, we reformulate the dataset by grouping every $n$ normal pair of observations together in a disjoint manner and treat them as a single data point in the training stage, which is denoted by $(\bX, \bY) \sim \cP_{0}^{\otimes n}$, where $\cP_{0}^{\otimes n}$ is the tensor product of measures and we utilize $\cD_0=\{\parent{\bX_i, \bY_i} \sim \cP_0^{\otimes n}\}_{i=1}^{N_0}$ to denote the collected dataset for training; similar notations also apply to other aforementioned data distributions. However, in practice, it is natural to relax the disjointness condition and instead use bootstrap subsampling without replacement to accumulate more training data. The core training objective function is given as follows and is minimized during the optimization process to ensure high canonical correlation scores between normal process sensing signals and normal quality data:\vspace{-0.03in}
\begin{align}
    \cL(\bff, \bg) = - \frac{1}{N_0}\sum_{(\bX, \bY) \in D_0} H_{n,p} \parent{\bff(\bX; \btheta_1), \bg(\bY; \btheta_2)},
    \vspace{-0.04in}
\end{align}
where $\bff(\cdot)$ and $\bg(\cdot)$ are neural network feature extractors to be trained, and the group size $n$ and hidden dimension $p$ are hyperparameters. In particular, the hidden correlation dimension is determined by iteratively training autoencoders on smaller subsets with progressively reduced hidden dimensions for both process signals and quality measurement data, until a desired reconstruction accuracy is achieved. This practical yet parsimonious principle ensures information preservation while minimizing redundancy. The procedure is illustrated in Algorithm \ref{alg: p-selection}, where $\epsilon_1$, $\epsilon_2 > 0$ are pre-specified parameters that denote the tolerance level of potential information loss and $p^*$ denotes the initial search dimension, which is typically chosen to be relatively large. It is important to note, however, that these parameters may vary significantly depending on the characteristics of different manufacturing processes and are typically determined using empirical domain knowledge or iterative refinement. Under this setting, the models are trained to extract non-linear representations of normal process signals and quality measurement data such that the canonical correlations between them are sufficiently high. 
% As a result, by leveraging the generalization power of neural networks, the model can distinguish between normal and abnormal process sensing signals by evaluating their correlation scores when paired with quality data collected from normal conditions.
\begin{algorithm}[t]
\caption{Correlation dimension selection\label{alg: p-selection}}
\begin{algorithmic}
\STATE \textbf{input:} $\cX_0, \cX_1, \cY_0, \cY_1, p^*, \epsilon_1, \epsilon_2$
\STATE Initialize $p=p^*+1$, $\cX = \{\cX_0, \cX_1\}$, and $\cY = \{\cY_0, \cY_1\}$.
\REPEAT
    \STATE $p \leftarrow p - 1$.
    \STATE Train an AutoEncoder $\texttt{AE}_1$ on $\cX$ with hidden dimension $p$ to convergence.
    \STATE Train an AutoEncoder $\texttt{AE}_2$ on $\cY$ with hidden dimension $p$ to convergence.
    \STATE Obtain reconstruction loss $\cL_1$ and $\cL_2$ of $\texttt{AE}_1$ and $\texttt{AE}_2$ on validation datasets.
\UNTIL{$\{\cL_1 > \epsilon_1\} \bigvee \{\cL_2 > \epsilon_2\}.$}
\RETURN $p$.
\end{algorithmic}
\end{algorithm}
\allowdisplaybreaks

\subsection{DCCA-based Online Quality Monitoring}\label{ssec: online-monitoring}
\begin{algorithm}[t]
\caption{Online correlation evaluation}
\label{alg: online-monitor}
\begin{algorithmic}
\STATE \textbf{input:} $\bX^*, \cX_0, \cY_0, d(\cdot, \cdot), \bff(\cdot; \btheta_1), \bg(\cdot; \btheta_2)$
\STATE Initialize $\bY^* \leftarrow [\ ]$.
\FOR{each $\bx_i$ in $\bX^*$}
    \STATE Find $\bx'=\argmin_{\bx \in \cX_0} d(\bx_i, \bx)$ and its corresponding quality matching $\by' \in \cY_0$.
    \STATE $\bY^* \leftarrow \bY^* \bigcup \ \{\by'\}$.
\ENDFOR
\STATE Compute $\rho^* = H_{n,p}(\bff(\bX^*; \btheta_1), \bg(\bY^*; \btheta_2))$.
\RETURN $\rho^*$
\end{algorithmic}
\vspace{-0.0in}
\end{algorithm}
In this section, we discuss how a pre-trained DCCA-based feature extraction model can be utilized to conduct online quality monitoring. The underlying idea is to apply the same principle used in the offline training scheme as outlined in Figure \ref{fig: high-level-intro}: when normal process signature signals are paired with the nearest referenced normal quality data from the historical dataset, their feature correlation scores should be high. In contrast, when abnormal process signature signals are paired with any normal quality data, their feature correlations are expected to have a lower mean and a higher variance, since no relationship between them was established during training. However, in real manufacturing processes, direct evaluation of canonical correlation scores between process signals and quality data is typically infeasible since only online process signals $\cX$ are observed. We overcome this obstacle by identifying surrogate normal quality data for canonical correlation computation, that is, searching for the nearest referenced target normal quality data $\bY^*$ for each incoming process signal window $\bX^*$ so that $(\bX^*, \bY^*)$ retains the correlation of the incoming data.

In this paper, the surrogate $\bY^*$ is calculated using a nearest neighbor search. The intuition is to find the normal quality data that is most likely to serve as the true counterpart for $\bX^*$ with the assumption that $\bX^*$ is a window of normal process signature signal. In this way, a strong canonical correlation between $\bX^*$ and $\bY^*$ can be obtained when $\bX^*$ is normal. On the other hand, when the incoming process signal $\bX^*$ is abnormal, pairing it with arbitrary quality data from normal processes results in a weaker canonical correlation score. Mathematically, the surrogate set $\bY^*$ is formulated as follows:\allowdisplaybreaks
\begin{align}
    \bY^* \coloneqq \Big\{\by'\Big| \bx'= &\argmin_{\bx \in \cX_0}d(\bx^*, \bx),(\bx', \by') \in (\cX_0, \cY_0), \forall \bx^* \in \bX^*\Big\}, \vspace{-0.08in}
\end{align}
where $d(\cdot, \cdot)$ is a pre-specified distance function. The full online correlation evaluation algorithm is given in Algorithm \ref{alg: online-monitor}.

Utilizing this online correlation evaluation method, we can define the following threshold-based detector $\eta: \reals^{d_1} \mapsto \{0,1\}$ on online process sensing signals. Given an offline-trained model and a suitable distance function $d(\cdot, \cdot)$, the monitoring function is given as follows:
% \vspace{-0.03in}
\begin{equation}
    \eta(\bX^*; \tau) = \mathbb{I}[\rho^*(\bX^*) < \tau],
    \vspace{-0.03in}
\end{equation}
where $\mathbb{I}(\cdot)$ is the indicator function, $\rho^*(\bX^*)$ denotes the total correlation score for $\bX^*$ obtained using Algorithm \ref{alg: online-monitor}, and $\tau\in [0,p]$ is the decision threshold determined by controlling the Type-I error rate of the method, that is, $\tau = \sup_{\tau \in [0,p]} \{\tau: \bbE_{\bX^* \sim \cP_{0,x}^{\otimes n}}[\eta(\bX^*; \tau)] \leq\alpha\}$, and empirically, this is conducted by controlling Type-I error rate on validation data under the normal process condition.

%% file: src/theory.tex
\section{Theoretical Guarantees}\label{sec: theory}
In this section, we present rigorous theoretical guarantees regarding the effectiveness of the proposed methods in empirical settings. For simplicity purposes, $\bff_{\btheta} \in \cF$ is used to denote the combined function of the two feature extraction networks $\bff(\cdot;{\btheta_1}),\bg(\cdot;{\btheta_2}) \in \cF$, parameterized by $\btheta$. For any $\bff_{\btheta} \in \cF$, let $\wt\cP_{0}({\bff_{\btheta}})$ be the corresponding feature distribution of normal process signature signals and quality measurement data ($\cP_{0}$) induced by the function $\bff_{\btheta}$. For $(\bu, \bv) \sim \wt\cP_{0}({\bff_{\btheta}})$, let $\rho^*_{0,\bff_{\btheta}}$ denote the true population canonical correlation coefficients between random vectors $(\bu, \bv)$ under the normal condition. 
With these notations, we represent the theoretical loss associated with $\bff_{\btheta}$ as $\bbE[\cL^*(\bff_{\btheta})] = -\rho^*_{0,\bff_{\btheta}}$ and the expected empirical loss as $\bbE[\cL(\bff_{\hat\btheta})]$, where $\bff_{\hat\btheta}$ is the empirical risk minimizer. 

\begin{figure}
    \centering
    \includegraphics[width=1.0\linewidth]{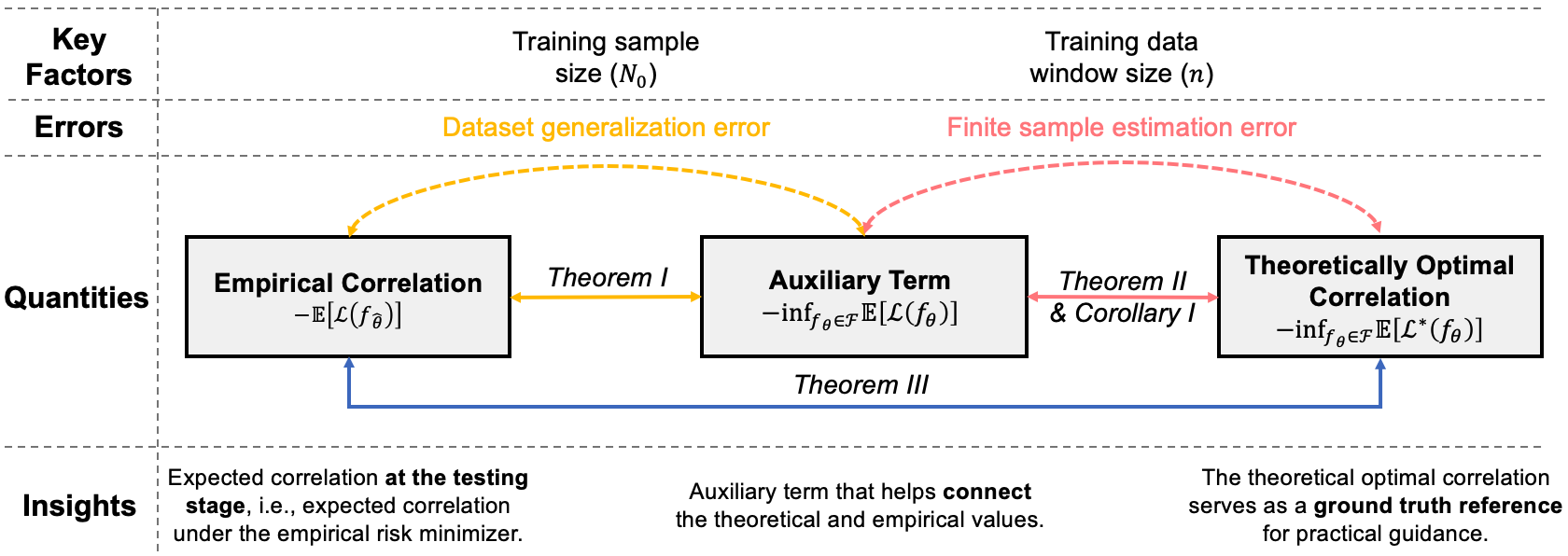}
    \caption{An overview of the theoretical guarantee for the proposed method. The correlation score of interest refers to the correlation between normal process signature signals and their corresponding normal quality data. This theoretical guarantee provides a solid foundation for assessing the practical utility of the proposed method in empirical settings.}
    \label{fig: theory}
\end{figure}

The objective is to establish a statistical learning bound that quantifies the deviation between $\bbE[\cL(\bff_{\hat\btheta})]$ and the theoretical optimum $\inf_{\bff_{\btheta} \in \cF} \bbE[\cL^*(\bff_{\btheta})]$, where the first quantity can be understood as the expected loss in the testing stage. The generalization error of the proposed loss function can be decomposed into two components: the error arising from the finiteness of the training dataset and the estimation error of the canonical correlation coefficient involved in the loss function. In particular, we investigate these two errors separately by connecting them with an intermediate term $\inf_{\bff_{\btheta} \in \cF}\bbE[\cL(\bff_{\btheta})]$ and then chaining them together to derive the desired bound. A visual illustration of two types of potential errors in the training stage and the theoretical analysis framework is given in Figure \ref{fig: theory}.

We first provide Theorem \ref{thm: rmc} to demonstrate that when the number of training samples is sufficiently large, the difference between $\bbE[\cL(\bff_{\hat\btheta})]$ and the auxiliary term $\inf_{\bff_{\btheta} \in \cF}\bbE[\cL(\bff_{\btheta})]$ approaches zero with high probability. Specifically, the target difference is bounded by two terms, each exhibiting a decay rate of $\mathcal{O}(N_0^{-1/2})$.
\begin{thm}\label{thm: rmc}
    Let $\bff_{\hat\btheta} \in \cF$ denote the empirical risk minimizer. Then under certain regularity conditions, for any $\delta > 0$, the following bound holds with probability at least $1 - \delta$,
    \begin{align}
        \bbE[\cL(\bff_{\hat\btheta})] & - \inf_{\bff_{\btheta} \in \cF}\bbE[\cL(\bff_{\btheta})]
         \leq 4L \cR_{N_0}(\cF)+ \sqrt{\frac{2p^2\log(1/\delta)}{N_0}},\vspace{-0.04in}
    \end{align}
    where $p$ is the total correlation dimension, $N_0$ denote the size of dataset $\cD_0$, $L > 0$ is the Lipschitz constant of $H_{n,p}(\cdot)$, and $\cR_{N}(\cF)$ is the Rademacher Complexity of function class $\cF$ \citep{bartlett2002rademacher}. 
\end{thm}
To account for the generalization error caused by the finite-sample estimation of canonical correlation coefficients utilized in the loss function, we derive the following non-asymptotic error bound for the canonical correlation estimation based on finite windows (Theorem \ref{thm: corr-nonasymptotic}). 
\begin{thm} (Non-asymptotic error bound for canonical correlation estimate) \label{thm: corr-nonasymptotic}
    Let $\rho^*$ denote the true $p$-dimensional canonical correlation coefficient between two random vectors $\bx$ and $\by$. Let $\{(\bx_i, \by_i)\}_{i \in [n]}$ be an i.i.d. finite $n$-sample estimate with $\norm{\bx_i}_2^2, \norm{\by_i}_2^2\leq B$ and $\bX, \bY$ be the collected data matrix, then with probability at least $1 - \delta$, 
    \begin{equation}
        \Big\lvert H_{n,p}(\bX, \bY) - \rho^* \Big\rvert \leq \frac{4Cp\log(6p / \delta)}{3n} + \sqrt{\frac{2C^2p^2\log (6p / \delta)}{n}},
    \end{equation}
    where $C$ is a constant function of $B, \lambda_{min}$ and $\lambda_{max}$, where $\lambda_{min}$ and $\lambda_{max}$ denote the range of eigenvalues of the covariance and cross-covariance matrix. 
\end{thm}
According to Theorem \ref{thm: corr-nonasymptotic}, it indicates that the estimation error decays at a rate of \(\mathcal{O}(n^{-1/2})\), implying that the gap shrinks quickly as the window size grows. For simplicity purposes, we employ $\xi_0(\delta;\bff_{\btheta})$ to denote the non-asymptotic error with confidence level $1 - \delta$ between the $n$-sample estimates and true correlations for $(\bu, \bv) \sim \wt\cP_{0}(\bff_{\btheta})$. The following Corollary \ref{cor: epsilon} connects the intermediate quantity $\inf_{\bff_{\btheta} \in \cF}\bbE[\cL(\bff_{\btheta})]$ and the theoretical optimum $\inf_{\bff_{\btheta} \in \cF} \bbE[\cL^*(\bff_{\btheta})]$, which helps us to quantify the overall error bounds.
\begin{cor} \label{cor: epsilon}
Under certain regularity conditions, for all $\delta > 0$, the following bound holds with probability at least $1 - \delta$,\allowdisplaybreaks
    \begin{equation}
        \Big\lvert \inf_{\bff_{\btheta} \in \cF}\bbE[\cL(\bff_{\btheta})] - \inf_{\bff_{\btheta} \in \cF} \bbE[\cL^*(\bff_{\btheta})]\Big\rvert \leq \xi_0^*(\delta),\vspace{-0.05in}
    \end{equation}
    where $\xi_0^*({\delta}) \sim \cO(n^{-1/2})$ is a function of $\delta$ constructed via the non-asymptotic error bounds $\xi_0(\delta;\bff_{\btheta})$ for $\bff_{\btheta} \in \cF$ and $\epsilon$-net techniques \citep{haussler1992decision}. The complete form and derivation of $\xi_0^*(\delta)$ is provided in Appendix A.4.
\end{cor}
Combining Theorem \ref{thm: rmc} and Corollary \ref{cor: epsilon}, we can derive the final statistical learning bound for the proposed objective function, which is presented in the following Theorem \ref{thm: final-bound}.
\begin{thm}\label{thm: final-bound}
Under the same setting described in Theorem \ref{thm: rmc} and Corollary \ref{cor: epsilon}, for any $\delta > 0$, the following holds with probability at least $1 - \delta$,
\begin{align}
     \bbE[\cL(\bff_{\hat\btheta})] \leq \inf_{\bff_{\btheta} \in \cF} &\bbE[\cL^*(\bff_{\btheta})] + \xi_0^{*}(\delta/2) + 4L \cR_{N_0}(\cF)+ \sqrt{\frac{2p^2\log(2/\delta)}{N_0}},
\end{align}
which provides an upper bound for the proposed loss function in the empirical setting. Furthermore, as $\xi_0^*(\delta/2) \sim \cO(n^{-1/2})$, it holds that asymptotically, the error terms will vanish as the number of training samples ($N_0$) and the number of samples within each window ($n$) used to estimate canonical correlations are sufficiently large.
\end{thm}
In practice, these statistical bounds provide a guideline for selecting parameters for the proposed method. Next, we provide a natural extension of our proposed framework to scenarios where limited abnormal process signals are available. Theoretically, we show that under certain separability assumptions, the gap in correlation scores between normal and abnormal samples can be empirically guaranteed, thereby validating the effectiveness of the proposed monitoring method. Specifically, the objective function can be extended to,
\vspace{-0.02in}
\begin{align}\label{eq: l_reg}
     \widetilde\cL(\bff, \bg) =& \ -\frac{1}{N_0}\sum_{(\bX, \bY) \in D_0} H_{n,p} \parent{\bff(\bX; \btheta_1), \bg(\bY; \btheta_2)} \notag\\
     + & \ \beta\frac{1}{N_1}\sum_{(\bX^*, \bY^*) \in D_1}H_{n,p} \parent{\bff(\bX^*; \btheta_1), \bg(\bY^*; \btheta_2)}
\end{align}
where $\beta \in [0,1]$ is a hyperparameter to be tuned that controls the magnitude of regularization. Here we slightly abuse the notation to use $\cP_1 = \cP_{1,x} \times \cP_{0,y}$ to represent the Cartesian product of two distributions, and similarly, $\cD_1 = \{\parent{\bX_i, \bY_i} \sim \cP_1^{\otimes n}\}_{i=1}^{N_1}$ denotes the abnormal dataset. The crucial assumption becomes that there exists a feature extractor $\bff_{\btheta^\circ} \in \cF$ such that the dependency or correlation scores calculated on the normal and abnormal datasets are distinct. This assumption is intuitive in the context of manufacturing applications since strong dependencies are expected between normal process and quality data. In contrast, for abnormal process signals, it is unreasonable to assume a non-trivial association with normal quality measurements.
\begin{assum}
(Existence of separable mapping) There exists $\bff_{\btheta^\circ} \in \cF$ such that $\rho^*_{0,\bff_{\btheta^\circ}} - \rho^*_{1,\bff_{\btheta^\circ}} \geq cp$, where $c \in [0,1]$ quantifies the gap of canonical correlations between data coming from the normal ($\cP_0$) and abnormal ($\cP_1$) distribution.
\end{assum}
Using the data separability assumption, we provide a similar upper bound for the extended objective function and conclude that the correlation gap is non-trivial when evaluating in an empirical setting, which is outlined in Theorem \ref{thm: final}.
\begin{thm}\label{thm: final}
Under the same setting as Theorem \ref{thm: rmc} but with the extended objective function, for any $\delta > 0$, the following holds with probability at least $1 - \delta$,
\begin{align}
     \bbE[\wt\cL(\bff_{\hat\btheta})] \leq -cp &+ \xi^{*}(\delta/2) + 4L\parent{\cR_{N_0}(\cF)+\beta\cR_{N_1}(\cF)}\notag\\
     &+ \sqrt{2\log(2/\delta)}p\parent{N_0^{-1/2} + \beta N_1^{-1/2}}, 
\end{align}
where $\xi^*(\delta / 2)$ is a function determined by the non-asymptotic errors derived in Theorem 2 and its complete form is provided in Appendix A.3. This Theorem provides an upper bound for the expectation of the modified objective with regularization terms. After decomposition, this states that the score gap between normal and abnormal distributions is preserved under the empirical minimizer $\bff_{\hat\btheta}$.
\end{thm} 
In particular, the decay rate of the last two terms in Theorem \ref{thm: final} follows \(\mathcal{O}((N_0 \wedge N_1)^{-1/2})\), while the second term also exhibits a decay rate of \(\mathcal{O}(n^{-1/2})\), as previously observed. This ensures that asymptotically, normal and abnormal samples can be distinguished by their correlation scores in the empirical setting when the model is trained on the extended loss function with available abnormal samples.

%% file: src/simulation.tex
\section{Simulation Study}\label{sec: simulation}
\begin{figure}
    \centering
    \includegraphics[width=1.0\linewidth]{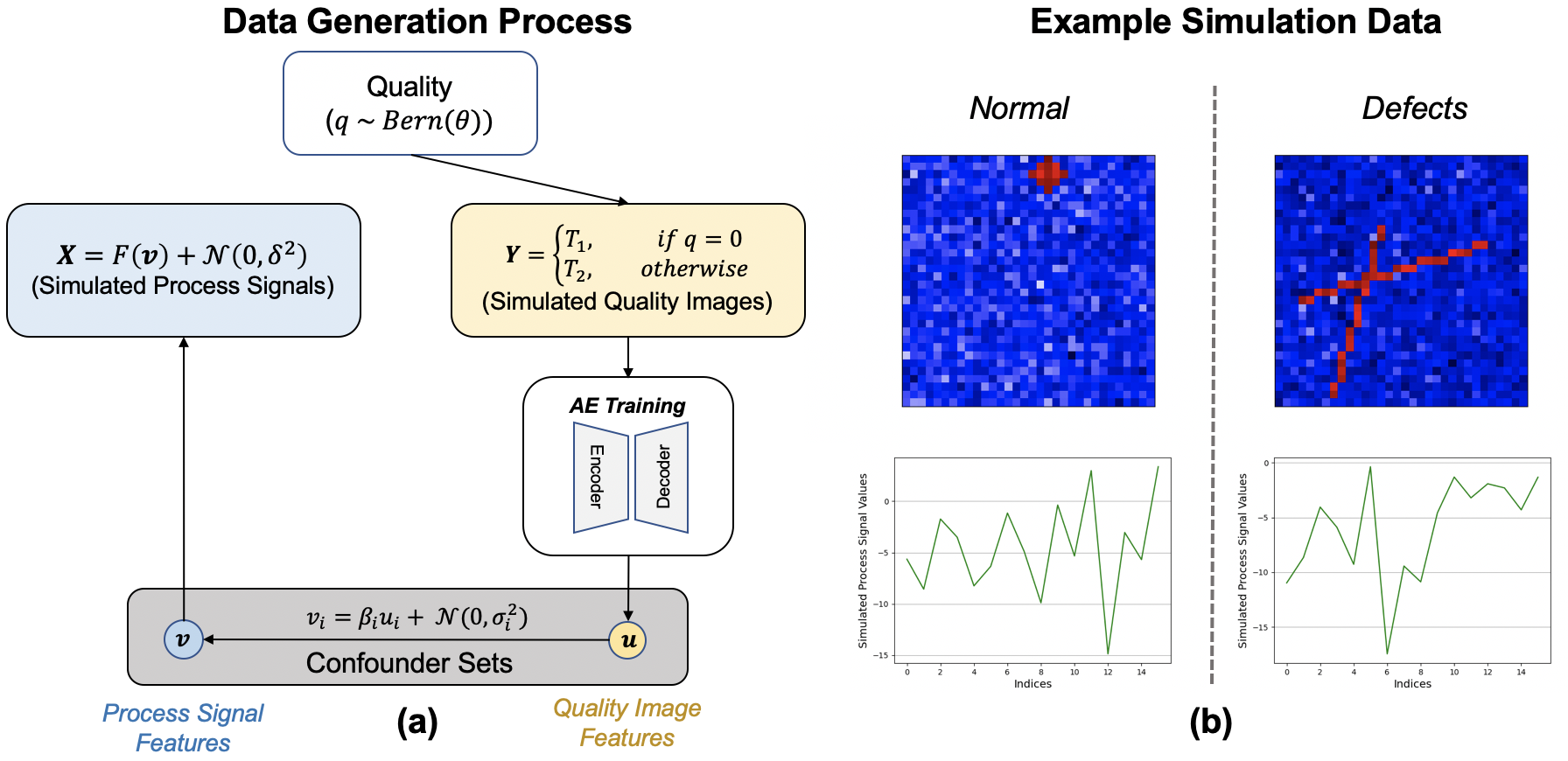}
    \caption{Illustration of the data generation process used in the simulation experiments (left) and a visualization of example simulation data (right).}
    \label{fig: simulation-dgp}
\end{figure}

In this section, we demonstrate the effectiveness of the proposed online quality monitoring approach through a simulation study. To imitate the real-world data flow, we generate the simulation data reversely starting from the quality label, where we sample $10000$ binary labels from $q_i \sim \text{Bern}(\theta)$ with $\theta$ being a pre-specified parameter controlling the expected fraction of non-conforming parts and $q_i=0$ representing normal quality. Quality images of normal and abnormal categories with dimension $d_0 \times d_0$ and binary pixel values are generated according to pre-specified heuristic functions $T_1 (\cdot)$ and $T_2(\cdot)$ introduced in Table \ref{tab: heuristics}. At a high level, this simulation study generates quality images with small pinholes as normal samples, while images with long cracks are treated as defects. This setup imitates a manufacturing process where pinholes are acceptable and do not affect product performance severely, whereas long cracks are considered critical defects. Figure \ref{fig: simulation-dgp} (b) provides corresponding example quality images for each category.
\begin{table}[t!]
\centering\small
\caption{Heuristic functions used for generating simulation data. When generating images, a $d_0 \times d_0$ blank image is provided as the background before any shapes are generated.}
\begin{tabular}{|>{\centering\arraybackslash}m{2.5cm}|>{\centering\arraybackslash}m{12.5cm}|}
\hline
Heuristics & Detailed Descriptions\\ \hline
$T_1(d_0, n, lb, ub)$ & \scriptsize The number of circles $m$ is first chosen uniformly from the set $[n]$. The radius and center of the circles are given by $\{r_i \sim U(lb, ub): i\in [m]\}$ and $\{c_i \sim U(r_i, d_0-r_i) \times U(r_i, d_0-r_i): i \in [m]\}$.\\ \hline
$T_2(d_0, n, t)$ & \scriptsize The number of cracks $m$ is first chosen uniformly from $[n]$. The starting and end locations (2D) are $\{s_i \sim U(0,d_0) \times U(0,d_0): i \in [m]\}$ and $\{e_i \sim U(0,d_0) \times U(0,d_0): i \in [m]\}$, respectively. The thickness of the line $\{t_i, i \in [m]\}$ is chosen uniformly from the set $[t]$. The cracks are then generated using \texttt{cv2.line()} function with the above parameters.\\ \hline
$F(\bv; h, h_1, h_2)$ & \scriptsize Define $\bM_1 \in \reals^{h_1 \times h}$ and $\bM_2 \in \reals^{h_2 \times h_1}$ with $M_1^{(ij)} \sim \cN(1, 1)$ and $M_2^{(ij)} \sim \cN(0, 1)$. Then, the process sensing signal is generated by $F(\bv) = \text{logit}(\lvert \bM_2 \ \text{ReLU}\parent{\bM_1\bv}\rvert) \in \reals^{h_2}$.\\ \hline
\end{tabular}
\label{tab: heuristics}
\end{table}

To ensure correspondence between the simulated quality images and spectrum-like process signals, an AutoEncoder \citep{vae} is utilized to extract features $\bu \in \reals^h$ for the image signals, and the corresponding features of process signature signals are generated by $\{\bv: v_i = \beta_iu_i + \cN(0,\sigma_i^2), \forall i \in [h]\}$, where $\{(\beta_i, \sigma_i)\}_{i=1}^h$ are chosen parameters that describe the true linear relationships between each dimension of the features of quality and process sensing signals. The process signals are then generated following the heuristic function $F(\cdot) + \cN(0,\delta^2)$ presented in Table \ref{tab: heuristics}. Intuitively, the magnitude of noise and the complexity of quality and process signals generation functions dictate the difficulty levels of model training and the experiments. Figure \ref{fig: simulation-dgp} (a) outlines this data generation process, and details on generation parameters can be found in Appendix B.1.
In general, this data generation process guarantees the existence of strong correlations between the generated quality and process signature signals, ensuring that the simulation dataset is well-aligned with real-world scenarios. In addition, the proposed data generation process can be treated as the reverse process of the real-world data flow, where $\{\bu, \bv\}$ represents the sets of unobserved confounder variables, and the established linear relationship between $\bu$ and $\bv$ enables us to consider them as a single entity in the analysis. 

% In particular, we set $h=6$, $\bbeta = [0.3, 1, 2, 0.4, -0.5, -0.7]$, and $\sigma_i=0.01$ for all $i \in [6]$ in the linear mapping within the unobserved confounder sets. The process signals are generated using $F(\bv; 6, 32, 16)$, while for the quality images, we investigate three distinct combinations of normal and abnormal generation heuristics: $T_{\text{cir}}(32, 3, 3, 6)$, $T_{\text{crk}}(32, 2, 2)$, and $T_{\text{cir}}'(32, 3, 3, 6, 0.3)$ as described in Table \ref{tab: heuristics}. These heuristics correspond to different difficulty levels for the monitoring tasks. Figure \ref{fig:simulation-eg} provides examples of generated process signals and quality data from each case. 
In all simulation experiments, the dimensions of the generated process signals and quality data are $\cX \in \mathbb{R}^{10000 \times 16}$ and $\cY \in \mathbb{R}^{10000 \times 1 \times 32 \times 32}$, respectively. For each configuration, we reserve 5000 randomly selected data points for testing, while the remaining data is divided evenly into training and validation sets at a ratio of 8:2. In this section, we inherit the notation from Section \ref{sec: method} and utilize $\cX_0, \cX_1, \cY_0$ and $\cY_1$ to represent the process signals and quality images of different labels in the training dataset.

\subsection{Offline Model Training}
We train the proposed model without the regularization term in Eq.(\ref{eq: l_reg}), assuming the absence of abnormal samples, where we organize the normal training data $\cX_0$ and $\cY_0$ into groups of size $n=25$. As introduced in Section \ref{sec: method}, we treat these 25 samples as a single data point and utilize them to estimate canonical correlations during the training stage. In this simulation study, we also relax the requirement that those grouped data points are disjoint, since this does not compromise the essential objective that normal pairs of process signals and quality data should demonstrate strong correlations after training. This relaxation also allows us to augment the training and validation dataset by generating more data. In particular, we generate 1000 data points for both training and validation data with group size $n=25$ from the normal training set $\cX_0$ and $\cY_0$ by bootstrap sampling without replacement. Once the training and validation datasets are constructed, the proposed DCCA-based model is trained according to Eq. \ref{eq: l_reg} using stochastic gradient descent, and the decision threshold is selected by controlling Type-I error rate to be $0.05$ on the validation dataset, respectively. Detailed introduction to data grouping operations, model architecture, and optimization parameters can be found in Appendix B.2.

\subsection{Online Benchmarking Experiments}
After offline training, we evaluate the proposed online monitoring framework on the holdout set in an online manner, where similarly, we utilize $\cX_0^*$ and $\cX_1^*$ to represent normal and abnormal process signature signals in the testing dataset, respectively. The quality images in the test sets remain unused as they can not be observed in an online setting. The first step in testing is to create a grouped dataset. Specifically, we construct the test dataset by sampling 2,000 data points of size 25 from the generated normal ($\cX_0^*$) and abnormal process signature signals ($\cX_1^*$), denoted by $\cD^*_0$ and $\cD_1^*$, respectively. We compare the proposed method against three baseline approaches, including traditional $T^2$ control charts using PCA and PLS features as monitoring statistics, respectively, and a naive classification-based method that monitors features of process signature signals. Since the classification-based method requires abnormal samples for training, we assume the availability of a limited set of 250 abnormal samples. All other methods are trained exclusively on normal process data. Appendix B.2 provides a detailed introduction to these baseline approaches.

\begin{figure}[t]
\begin{minipage}{.33\textwidth}
\centering
    \includegraphics[scale=0.36]{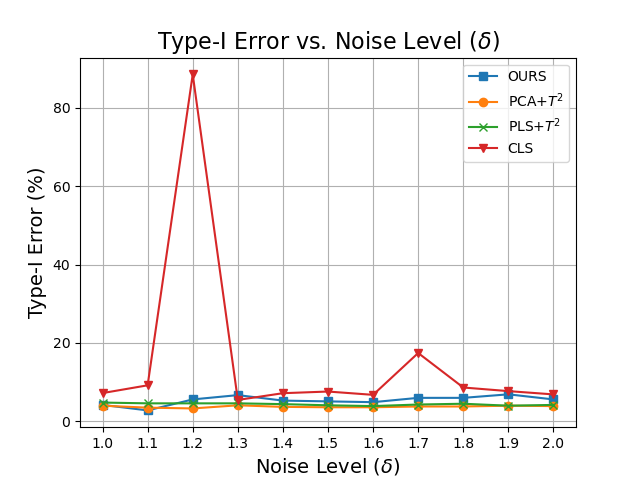}
    \captionsetup{justification=centering}
\end{minipage}
\begin{minipage}{.33\textwidth}
\centering
    \includegraphics[scale=0.36]{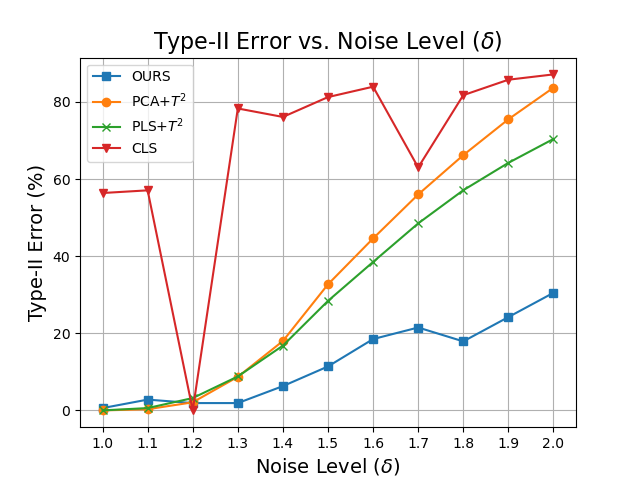}
    \captionsetup{justification=centering}
\end{minipage}
\begin{minipage}{.3\textwidth}
\centering
    \includegraphics[scale=0.36]{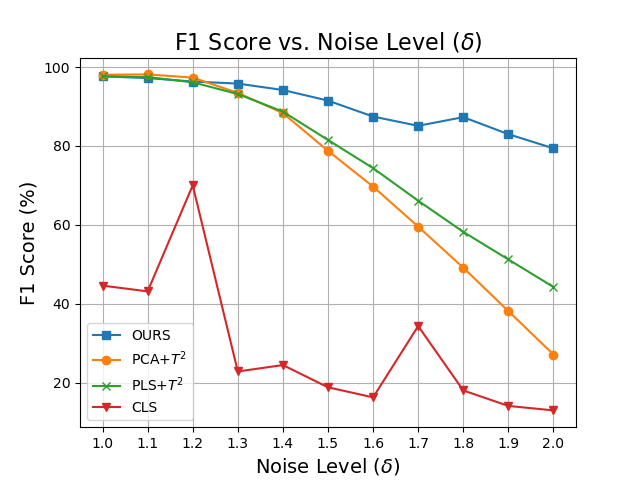}\captionsetup{justification=centering}
\end{minipage}
\caption{Simulation experiment results for the proposed and baseline methods across various noise levels ($\delta$) introduced during the data generation process. \label{fig: simulation-results}}
\end{figure}

In the simulation experiments, we follow the aforementioned data generation procedure and introduce varying noise levels to simulate different levels of difficulty. Specifically, we perturb the process signature signals with Gaussian noise having standard deviations ranging from 1 to 2, while the overall signal magnitude remains approximately within the range of 0 to 20. Figure \ref{fig: simulation-results} presents the results under varying experimental configurations. Under low noise levels, the proposed method exhibits comparable performance to traditional control chart-based approaches across all evaluation metrics. However, as the level of noise and difficulty increases, it demonstrates substantial improvements, particularly in reducing Type-II error rate and enhancing the overall F1 score. This performance gain can be attributed to the limitations of variation-based methods, which are inherently designed to capture the variations in process signature signals and are therefore more susceptible to noise. In contrast, the proposed method is explicitly trained to extract quality-relevant structures, making it more robust to noise-induced variations in tasks related to quality identification. In addition, the classification-based benchmark exhibits a notable performance degradation primarily due to the limited number of defective training samples, which is a common constraint in real-world manufacturing applications. Complete discussions of the simulation study are provided in Appendix B.

%% file: src/case.tex
\section{Case Study}\label{sec: case-study}
\subsection{Data Collection \& Processing}
\begin{figure}[t]
    \centering
\includegraphics[width=1.0\linewidth]{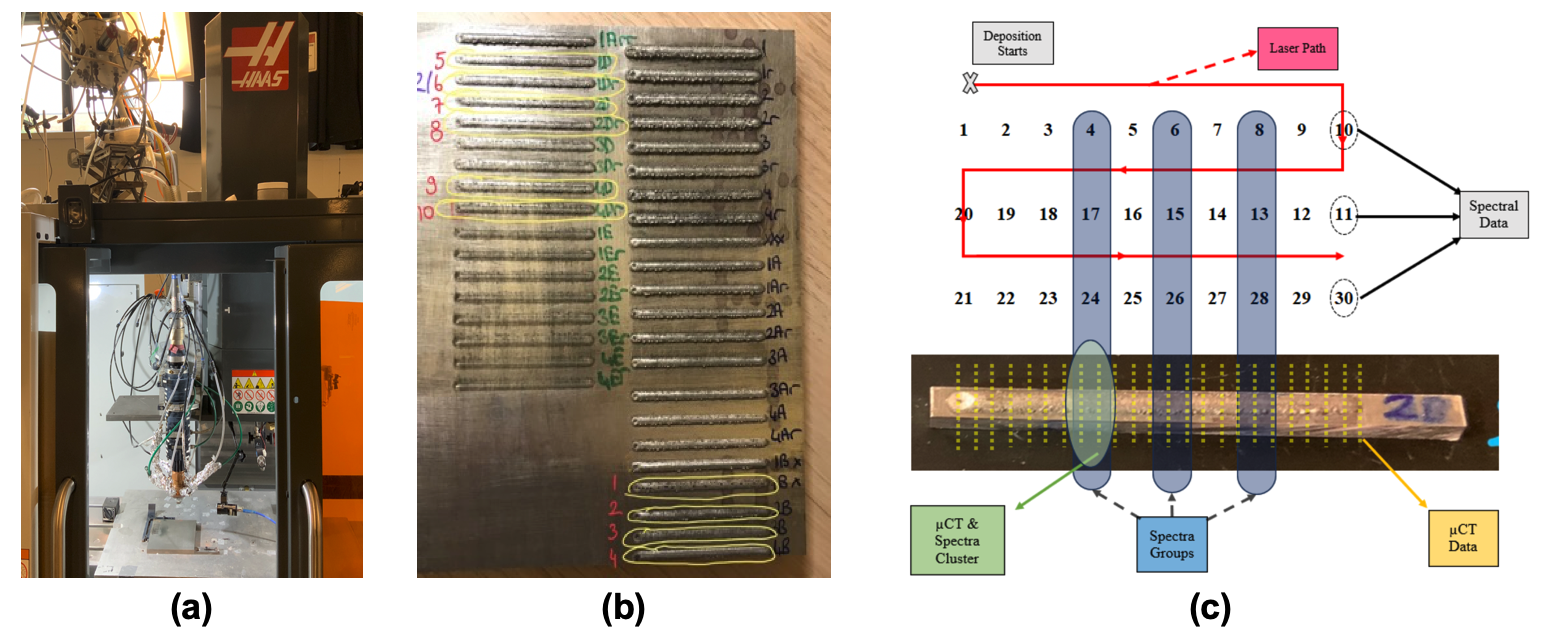}
    \caption{An overview of the DMD case study experimental setup is shown in (a), along with example finished parts in (b), and an illustration of the procedures for collecting in-situ spectra and offline CT scans in (c). In short, the printed part is a single line as shown in (b), and the laser path is represented by the red line in (c). For illustrative purposes, the spectra data are labeled from 1 to 30. Considering the deposition sequence, the correct grouping of spectra corresponds to positions such as 1–20–21, 2–19–22, and so on. \label{fig: example-alignment}}
\end{figure}

We conducted a case study on a Direct Metal Deposition (DMD) additive manufacturing process, collecting both in-situ optical emission spectra and offline quality CT scans. The experiments were performed using Aluminum 7075 alloy under two distinct process conditions, as detailed in Table \ref{tab: real-data-description}, to produce both normal and defective parts. Figure \ref{fig: example-alignment} provides examples of the representative finished parts and the corresponding data collection procedure. Based on positions, these online process signature signals and quality CT scans are aligned together. The objective is to leverage the easily obtained online optical emission spectra signals to identify low quality part whose CT scans show abnormal pattern yet are difficult to collect during the manufacturing process. A detailed description of the DMD experiment environment configuration is provided in Appendix C.1.

\begin{figure}[t]
    \centering
    \includegraphics[width=0.95\linewidth]{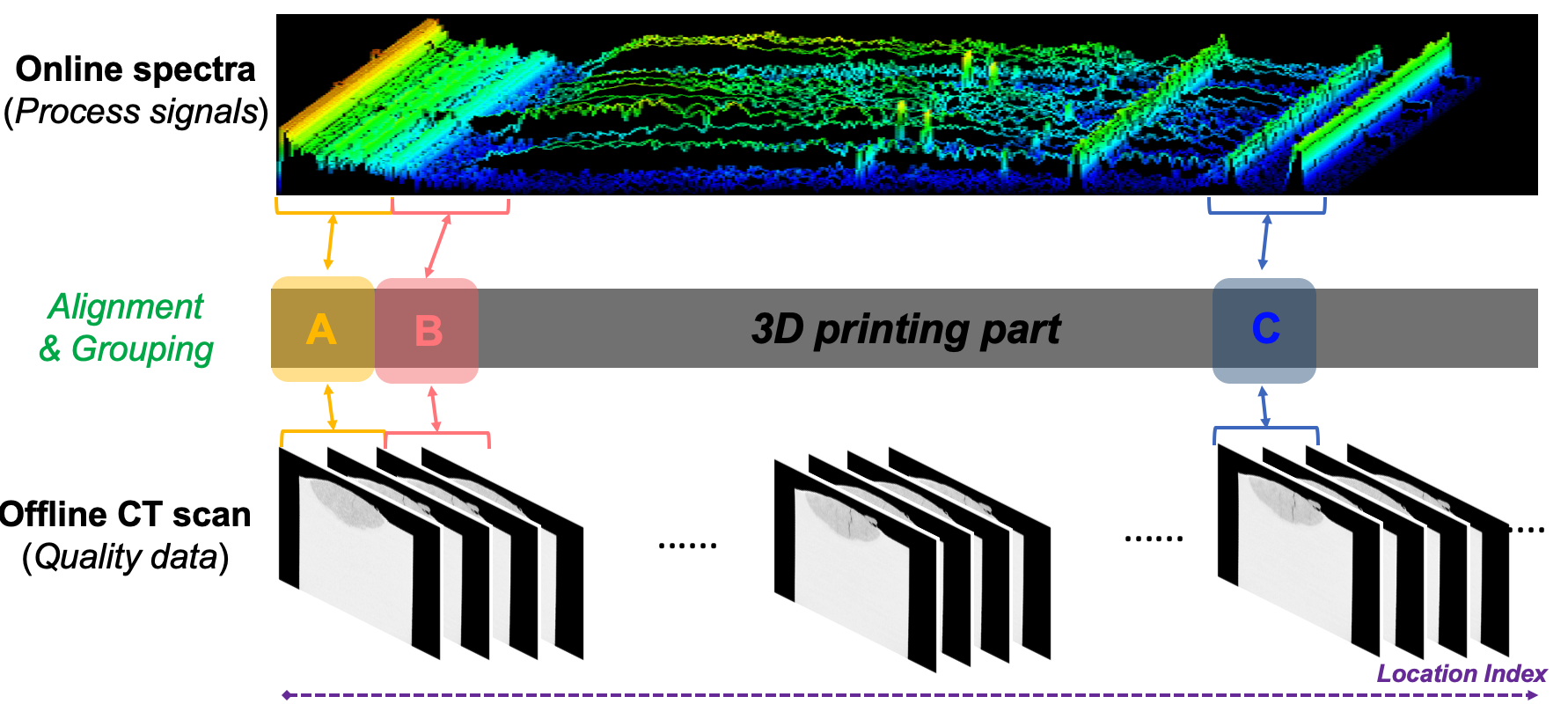}
    \caption{Data alignment process illustration and examples. At each printing location, corresponding online spectral process signals and offline CT quality scans are collected and aligned based on their respective sampling frequencies. For clarity of visualization, the spectra shown in this figure are illustrative rather than real data.}
    \label{fig: case-study-processing}
\end{figure}
After collecting raw data under various process conditions, the in-situ spectral sensing signals and offline CT scan quality images are processed and aligned based on their respective sampling frequencies at corresponding printing locations. Figure \ref{fig: example-alignment} (c) presents a detailed illustration of the alignment procedure, while Figure \ref{fig: case-study-processing} offers an abstract visualization of the process. Note that in simulation experiments, we assume an ideal setting with equal sampling frequencies for process signals and quality measurements to enable a one-to-one mapping between them. In real-world experiments, however, such exact correspondence and alignment are often infeasible due to differing sampling rates for online process signature signals and offline quality measurements. To address this, we align every five spectral sensing signals with 22 CT scan images, resulting in data pairs of $\bx \in \mathbb{R}^{5 \times 32}$ and $\by \in \{0,1\}^{22 \times 250 \times 730}$. For instance, in Figure \ref{fig: case-study-processing}, samples A, B, and C from different printing locations are obtained using the alignment process described in the same figure, and the resulting samples conform to the specified data pair dimensions.

These complex structures observed in this real additive manufacturing process further highlight the need for tensor representations and effective feature extraction methods for them. To simplify the feature extraction process, we further compress the window of five quality images by averaging their values and conducting threshold-based denoising, which effectively preserves defect-related information. This renders a binary quality image of dimension $250 \times 730$. More details on dataset alignment and preprocessing are discussed in Appendix C.2.
\begin{table}[t]
\setlength{\aboverulesep}{0pt}
\setlength{\belowrulesep}{0pt}
  \centering
  % \small
    \caption{Process parameters and description on collected spectra and CT scan dataset. \label{tab: real-data-description}}
  \scalebox{0.63}{
    \begin{tabular}{c|ccc|cc|cc|cc}
    \toprule
    \multirow{2}[6]{*}{\tb{Processes}} &\multicolumn{3}{c}{\tb{Process Configuration}} & \multicolumn{2}{c}{\tb{Data Dimension}} & \multicolumn{2}{c}{\tb{Data Dimension (processed)}} & \multicolumn{2}{c}{\tb{Aligned Dataset}}\\\cmidrule{2-10}
    & \makecell{Power \vspace{-0.2cm}\\(W)} & \makecell{Speed \vspace{-0.2cm}\\(mm/s)} & \makecell{Diameter \vspace{-0.2cm}\\(mm)} & \makecell{Spectra\vspace{-0.2cm}\\($L \times \lambda$)} & \makecell{CT scans\vspace{-0.2cm}\\($L \times C \times H \times W$)} & \makecell{Spectra\vspace{-0.2cm}\\($L \times \lambda$)} & \makecell{CT scans\vspace{-0.2cm}\\($L \times C \times H \times W$)} & \makecell{Training\vspace{-0.2cm}\\size} & \makecell{Testing\vspace{-0.2cm}\\size}\\\hline
    Normal & 1750 & 21 & 1 & $6759 \times 2038$ & $4751 \times 1 \times 674 \times 1000$ & $681 \times 32$ & $4751 \times 1 \times 250 \times 730$ & 89 & 40\\
    Abnormal & 1800 & 15 & 1 & $7376 \times 2038$ & $4046 \times 1 \times 674 \times 1000$ & $942 \times 32$ & $4046 \times 1 \times 250 \times 730$ & 135 & 40\\
    \bottomrule
    \end{tabular}
    }
\end{table}

\subsection{Process Shift Detection Experiment}
We conduct a process shift detection experiment to demonstrate the effectiveness of the proposed method. In this experiment, the proposed correlation-based model is trained using data from the known normal process condition and is tasked with distinguishing spectra sensing signals collected from the abnormal processes with different configurations. A detailed description of the process configuration and aligned dataset is provided in Table \ref{tab: real-data-description}. As for the model architecture, we utilize the LSTM-based encoder model proposed by \citet{sun2022situ}, followed by layers of convolutional and linear modules for spectra sensing signals. For CT scan data, we just employ a multi-layer convolutional neural network (CNN) \citep{lecun2015deep}. The hidden feature dimension is chosen to be $p=6$ for the case study experiments following the proposed parsimonious principle. 

For offline training, the dataset is split into training and validation sets in an 8:2 ratio. The two feature extraction models for spectra and CT scans are then trained under the DCCA-based framework using a group size of $n = 8$ for estimation. After training, the decision threshold is determined using a validation set consisting of 200 data points from the normal process, each with a common window size of 8. Due to the limited size of the validation dataset, these windowed samples are generated via bootstrap sampling without replacement. The threshold is chosen to control Type-I error rate at 0.1 on the validation set, based on the proposed Algorithm \ref{alg: online-monitor} involving matching and correlation evaluation. Similarly, a test set is constructed with 200 normal and 200 abnormal data points, each with a common group size of 8. The model is then evaluated on its ability to distinguish whether a given spectral window originates from a normal or abnormal process. Detailed model architectures and training parameters are provided in Appendix C.3.
\begin{figure}[t]
\centering
\begin{minipage}{0.39\textwidth}
    \centering
    \includegraphics[width=\textwidth]{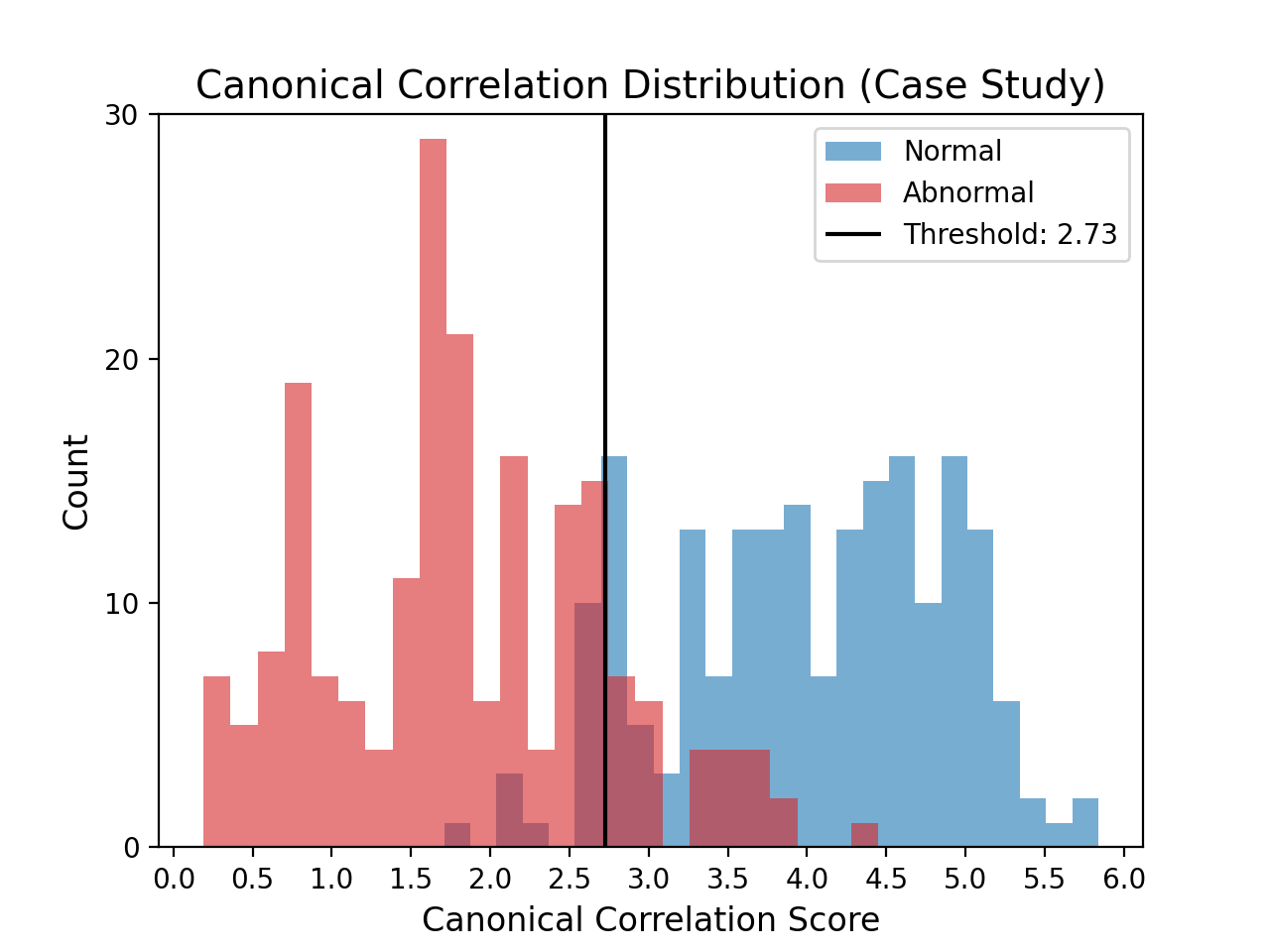} 
    % \caption{Caption for the figure}
    % \label{fig:example}
\end{minipage}%
\hfill
\begin{minipage}{0.60\textwidth}
    \centering
    \resizebox{\textwidth}{!}{%
    \begin{tabular}{|c|c|c|c|c|c|c|}
        \hline
        \multirow{2}{*}{\tb{Methods}} & \multicolumn{3}{c|}{Validation-based Threshold} & \multicolumn{3}{c|}{$\alpha$-Adjusted Threshold} \\\cline{2-7}
        & \tb{FPR$\downarrow$} & \tb{FNR$\downarrow$} & \tb{F1$\uparrow$} & \tb{FPR$\downarrow$} & \tb{FNR$\downarrow$} & \tb{F1$\uparrow$} \\\hline
        \texttt{PCA + $T^2$} & 63.5 & 4.0 & 73.75 & \tb{10.0} & 83.5 & 26.09 \\ \hline
        \texttt{PLS + $T^2$} & 74.0 & 5.5 & 74.39 & \tb{10.0} & 77.5 & 33.97 \\\hline
        % \texttt{DCCA + $T^2$} & 54.49 & \textbf{0.0} & 77.97 \\\hline
        \texttt{DCCA + NN} & \textbf{9.5} & 14.0 &  \textbf{87.98} & \tb{10.0} & \tb{14.0} & \tb{87.76} \\\hline
    \end{tabular}
    }
\end{minipage}
\caption{Histograms of canonical correlation scores with thresholds selected using the validation set for case studies (left) and corresponding numerical results (right). Results are reported using thresholds determined by our proposed Algorithm \ref{alg: online-monitor} on the validation dataset, as well as thresholds adjusted to control Type-I error rate across different methods.\label{fig: case-study-results}}
\end{figure}

Figure \ref{fig: case-study-results} presents the experimental results, where the histogram clearly illustrates a distinct separation in the canonical correlation scores between normal and abnormal process sensing signals. This separation validates the effectiveness of the proposed offline training strategy. Moreover, the proposed method achieves a strong performance in terms of Type-I error rate and F1 score, validating its good performance in practical in-situ monitoring scenarios. We further benchmark our approach against several existing feature extraction and monitoring techniques using $T^2$ statistics. While some baselines achieve lower Type-II error rates, their significantly higher Type-I error rates reveal their poor performance in effectively distinguishing between normal and abnormal process signals. Indeed, these methods tend to misclassify all incoming signals as defective, which indicates their limitations in capturing meaningful features from complex signals. Furthermore, it can be observed that although the decision thresholds are selected using the validation set with Type-I error rates controlled at 0.1, the actual Type-I error rates on the test set are significantly higher in those baseline methods, contrary to the expectation that they would remain close to 0.1. This discrepancy arises due to a strong violation of the normality assumption, which invalidates the usage of the $T^2$ statistic and leads to more random and unpredictable behavior. We also report results under a fixed Type-I error rate of 0.1 across all methods and observe that the proposed approach consistently outperforms the baselines by achieving a significantly lower mis-detection rate.
% Results for different fixed levels of the Type-I error rate are provided in Appendix E. 
Overall, these findings highlight the robustness and discriminative power of the proposed method for feature extraction in real-world applications

%% file: src/conclusion.tex
\section{Conclusion}
\label{sec: conclusion}
This paper extends the existing Deep Canonical Correlation Analysis (DCCA) network to effectively extract features from high-dimensional process signature signals and quality data. By ensuring that the extracted features from both modalities are well-correlated, the proposed model are combined with a nearest neighbor-based online monitoring algorithm to conduct efficient in-situ quality monitoring. The proposed method is inherently model-free, allowing it to address the challenges associated with data processing through the integration of different network architectures. It overcomes the limitations of traditional quality-oriented supervised feature extraction methods, which require quality labels or specifications and are therefore restricted to specific applications. While this work primarily focuses on improving in-situ process monitoring performance, we believe that this research lies a solid foundation for future works to explore leveraging the extracted correlated features to directly reconstruct quality-related images from process signature signals, potentially through the joint development of a generative model. We envision that such a future work could significantly enhance the interpretability of process monitoring decisions and enable more direct analysis and visualization of emerging defects.

\section{Data Availability Statement}
This paper presents both simulation experiments and a case study based on a real-world metal additive manufacturing dataset to demonstrate the effectiveness of the proposed method. The simulation data is provided to support reproducibility of the results. Due to licensing restrictions, the dataset used in the case study is not intended to make publicly available. However, case data access permission can be requested via emails to authors for appropriate use.

%% file: src/appendix.tex
\newpage
\appendix
\setcounter{lem}{0}
\setcounter{remark}{0}
\setcounter{cor}{0}
\setcounter{defn}{0}
\renewcommand{\thethm}{\Alph{section}.\arabic{thm}}
\renewcommand{\thelem}{\Alph{section}.\arabic{lem}}
\renewcommand{\thecor}{\Alph{section}.\arabic{cor}}
\renewcommand{\thedefn}{\Alph{section}.\arabic{defn}}
\renewcommand{\thefigure}{\Alph{section}.\arabic{figure}}
\renewcommand{\thetable}{\Alph{section}.\arabic{table}}

\input{src/appendix/proof}

\input{src/appendix/simulation}

\input{src/appendix/case}

%% file: src/appendix/proof.tex
\section{Proofs of Theoretical Results}
Let $\bZ = (\bX \ \bY) \in \reals^{2p \times n}$ denote the input to the canonical correlation function, where $\bX, \bY \in \reals^{p \times n}$ are essentially two views of the same data source. For instance, in this paper, $\bX$ and $\bY$ represent online process sensing spectra signals and offline CT scans, respectively. In addition, it is natural to assume that $n \gg p$, that is, the sample size is much greater than the feature dimension. Here we define the following functions for later usage:
\begin{align}
    &t_1(\bZ)=\bZ\bA\bZ^\top=\begin{bmatrix}
        \bX\bX^\top & \mathbf{0}_n\\
        \mathbf{0}_n & \mathbf{0}_n
    \end{bmatrix}, \bA = \begin{bmatrix}
        \bI_n & \mathbf{0}_n\\
        \mathbf{0}_n & \mathbf{0}_n
    \end{bmatrix},\\
    &t_2(\bZ) = \bZ\bB\bZ^\top=\begin{bmatrix}
        \mathbf{0}_n & \mathbf{0}_n\\
        \mathbf{0}_n & \bY\bY^\top
    \end{bmatrix},  \bB = \begin{bmatrix}
    \mathbf{0}_n & \mathbf{0}_n\\
    \mathbf{0}_n & \bI_n
    \end{bmatrix},\\
    &t_3(\bZ)= \bZ\bC\bZ^\top =\begin{bmatrix}
        \mathbf{0}_n & \bX\bY^\top \\
        \mathbf{0}_n & \mathbf{0}_n
    \end{bmatrix},    \bC = \begin{bmatrix}
    \mathbf{0}_n & \bI_n\\
    \mathbf{0}_n & \mathbf{0}_n
    \end{bmatrix}.
\end{align}
With these definitions, we can rewrite the canonical correlation function $H: \reals^{2p \times n} \mapsto \reals$ into the following form, which, in essence, is the closed-form solution for the maximization problem in canonical correlation analysis:
\begin{align}
    H(\bZ) &= \norm{\hat\bSigma_{11}^{-1/2}\hat\bSigma_{12}\hat\bSigma_{22}^{-1/2}}_{*}\notag\\
    &= \norm{\parent{\frac{1}{n-1}\bX\bX^\top}^{-1/2} \parent{\frac{1}{n-1}\bX\bY^\top}\parent{\frac{1}{n-1}\bY\bY^\top}^{-1/2}}_{*}\notag\\
    &= \norm{\parent{\bX\bX^\top}^{-1/2} \parent{\bX\bY^\top}\parent{\bY\bY^\top}^{-1/2}}_{*}\notag\\
    &=\norm{\parent{t_1(\bZ)_{[1:n, 1:n]}}^{-1/2} t_3(\bZ)_{[1:n, n+1:2n]}\parent{t_2(\bZ)_{[n+1:2n, n+1:2n]}}^{-1/2}}_*,\label{eq:corr-rewrite}
\end{align}
where $\norm{\cdot}_*$ denotes the matrix trace norm and $\bA_{[r_1:r_2, c_1:c_2]}$ is the submatrix operation of $\bA$. For instance, $t_1(\bZ)_{[1:n, 1:n]}$ denotes the upper left $n \times n$ block in matrix $t_1(\bZ)$.

\subsection{Lipschitz Continuity of $H_{n,p}(\cdot)$}
\begin{lem}\label{lem:lipschitz}
    Let $\bZ = (\bX, \bY) \in \reals^{2p \times n}$ and let $H_{n,p}: \chi_{\bZ} \mapsto \reals$ be the function that computes the $p$-dimensional canonical correlation using a finite sample $\bX$ and $\bY$ of size $n$. Then, $H_{n,p}$ is Lipschitz continuous when it is defined on the domain with $\sigma_{min} \leq \sigma (\bX), \sigma(\bY) \leq \sigma_{max}$, that is, the singular values of $\bX, \bY$ are bounded. 
\end{lem}

To prove this lemma, we need to analyze the Lipschitz continuity of each function that appears in Eq. \ref{eq:corr-rewrite} under the given regularity conditions. First, recall the following two useful facts from elementary linear algebra for $\bA \in \reals^{m \times n}$ with rank $r$: (1) $\norm{\bA}_2 \leq \norm{\bA}_F \leq \sqrt{r}\norm{\bA}_2$ and (2) $\norm{\bA}_F \leq \norm{\bA}_* \leq \sqrt{r}\norm{\bA}_F$.
\begin{remark}
    It is noteworthy that the imposed assumption on bounded singular values on data matrix $\bX$ and $\bY$ essentially implies the full row-rankness on $\bX$ and $\bY$, and hence positive-definiteness of $\bX\bX^\top$ and $\bY\bY^\top$. In the proposed method, this is a natural assumption as we expect the extracted features to be parsimonious, indicating no redundancy.
\end{remark}
\begin{lem}
    Given those regularity conditions, $t_1(\cdot), t_2(\cdot),$ and $t_3(\cdot)$ are Lipschitz continuous with respect to its input.
\end{lem}
\begin{proof}\allowdisplaybreaks
    Without losing of generality, we provide the proof for $t_1: \bZ \mapsto \bZ\bA\bZ^\top$. Let $\bZ_0, \bZ_1 \in \reals^{p \times 2n}$ be two arbitrary centered data matrices that satisfy the regularity conditions, then,
    \begin{align*}
        \norm{t_1(\bZ_0) - t_1(\bZ_1)}_2 &= \norm{\bZ_0\bA\bZ_0^\top - \bZ_1\bA\bZ_1^\top}_2\\
        &=\norm{\bZ_0\bA\bZ_0^\top - \bZ_1\bA\bZ_0^\top + \bZ_1\bA\bZ_0^\top - \bZ_1\bA\bZ_1^\top}_2\\
        &= \norm{(\bZ_0-\bZ_1)\bA\bZ_0^\top + \bZ_1\bA\parent{\bZ_0 - \bZ_1}^\top}_2\\
        &\leq \norm{(\bZ_0-\bZ_1)\bA\bZ_0^\top}_2 + \norm{\bZ_1\bA\parent{\bZ_0 - \bZ_1}^\top}_2\\
        &\leq \parent{\norm{\bA\bZ_0^\top}_2 + \norm{\bZ_1\bA}_2}\norm{\bZ_0-\bZ_1}_2\\
        &= \parent{\norm{\bX_0}_2 + \norm{\bX_1}_2}\norm{\bZ_0-\bZ_1}_2\\
        &\leq 2\max\{\sigma(\bX_0), \sigma(\bX_1)\}\norm{\bZ_0-\bZ_1}_2
    \end{align*}
    The proof for $t_2(\cdot)$ and $t_3(\cdot)$ follows from exactly the same structure but differs at the third last step. To summarize, we have that,
    \begin{align}
        \norm{t_1(\bZ_0) - t_1(\bZ_1)}_2 &\leq 2\max\{\sigma(\bX_0), \sigma(\bX_1)\}\norm{\bZ_0-\bZ_1}_2\leq 2\sigma_{max}\norm{\bZ_0-\bZ_1}_2,\\
        \norm{t_2(\bZ_0) - t_2(\bZ_1)}_2 &\leq 2\max\{\sigma(\bY_0), \sigma(\bY_1)\}\norm{\bZ_0-\bZ_1}_2\leq 2\sigma_{max}\norm{\bZ_0-\bZ_1}_2,\\
        \norm{t_3(\bZ_0) - t_3(\bZ_1)}_2 &\leq \parent{\max\{\sigma(\bX_0)\} + \max\{\sigma(\bY_1)\}}\norm{\bZ_0-\bZ_1}_2\leq 2\sigma_{max}\norm{\bZ_0-\bZ_1}_2
    \end{align}
    which shows the Lipschitz continuity of these three functions, and this completes the proof.
\end{proof}

\begin{lem}\label{lem: lipschitz-inv-sqrt} (Lipschitz Continuity for Inverse Square Root)
    Let $\bM \in \bS_{++}^p$ be a positive-definite matrix whose singular values are bounded below and above by $\sigma_M^{min}$ and $\sigma_M^{max}$, respectively. Then, in this domain, the matrix inversion function $f(\bM) = \bM^{-1}$ and the square root function $g(\bM) = \bM^{1/2}$ are Lipschitz continuous. Hence, the function $f \circ g(\bM)=\bM^{-1/2}$ is also Lipschitz continuous with respect to $\bM$.
\end{lem}
\begin{proof}\allowdisplaybreaks
    Given any two matrices $\bM_0, \bM_1 \in \bS_{++}^p$, we can obtain that,
    \begin{align*}
        \norm{\bM_0^{-1} - \bM_1^{-1}}_2 &= \norm{\bM_0^{-1}(\bM_1 - \bM_0)\bM_1^{-1}}_2\\
        &\leq \norm{\bM_0^{-1}}_2\norm{\bM_1^{-1}}_2\norm{\bM_1 - \bM_0}_2\\
        % &\leq p\norm{\bM^{-1}}_2\norm{\bM'^{-1}}_2\norm{\bM' - \bM}_F\\
        &\leq (1/\sigma_M^{min})^2\norm{\bM_0 - \bM_1}_2,
    \end{align*}
    which proves the Lipschitz continuity of matrix inversion in this specific domain. Similarly, we prove the Lipschitz continuity for the matrix square root function. For any $\bM_0, \bM_1 \in \bS_{++}^p$, we can obtain that,
    % \begin{align}
    %     \bQ^{-1}(\bM - \bM')\bQ &= \bQ^{-1}\parent{\sqrt\bM - \sqrt{\bM'}}\parent{\sqrt\bM + \sqrt{\bM'}}\bQ\\
    %     &= \bLambda\bQ^{-1}\parent{\sqrt\bM + \sqrt{\bM'}}\bQ.
    % \end{align}
    % After some simple algebra, we have that,
    % \begin{equation}
    %     \bLambda = \bQ^{-1}(\bM - \bM')(\sqrt{\bM} + \sqrt{\bM'})^{-1}\bQ.
    % \end{equation}
    \begin{align*}
        \norm{\sqrt\bM_0 - \sqrt{\bM_1}}_2 &= \norm{(\bM_0 - \bM_1)(\sqrt{\bM_0} + \sqrt{\bM_1})^{-1}}_2 \\
        &\leq \norm{(\sqrt{\bM_0} + \sqrt{\bM_1})^{-1}}_2\norm{\bM_0 - \bM_1}_2\\
        % &\leq \sqrt{\rank{\parent{(\sqrt{\bM} + \sqrt{\bM'})^{-1}}}}\norm{(\sqrt{\bM} + \sqrt{\bM'})^{-1}}_2\norm{\bM - \bM'}_F\\
        &\leq \sigma_{min}(\sqrt{\bM_0} + \sqrt{\bM_1})^{-1}\norm{\bM_0 - \bM_1}_2\\
        &\leq \parent{\sigma_{min}(\sqrt{\bM_0}) + \sigma_{min}(\sqrt{\bM_1})}^{-1}\norm{\bM_0 - \bM_1}_2\\
        &\leq \frac{1}{2\sqrt{\sigma_M^{min}}}\norm{\bM_0 - \bM_1}_2,
    \end{align*}
    where the above derivation utilizes the Weyl's inequality \citep{weyl1939} and the equivalence of singular values and eigenvalues of positive-definite matrices. This completes the proof of Lipschitz continuity of matrix square root function. Now, utilizing the fact that composition of two Lipschitz functions is also Lipschitz, we know that $f \circ g$ is also Lipschitz with respect to its input. Mathematically, we have that,
    \begin{equation}
        \norm{\bM_0^{-1/2} - \bM_1^{-1/2}}_2 = \norm{f \circ g(\bM_0) - f \circ g (\bM_1)}_2 \leq \frac{1}{2 {\sigma_{M}^{min}}^{3/2}} \norm{\bM_0 - \bM_1}_2,
    \end{equation}
    where the constant is obtained by multiplying the Lipschitz constant for $f$ and $g$ together.
\end{proof}

\begin{lem}
    The nuclear norm $\norm{\cdot}_*$ (matrix trace norm) is Lipschitz continuous.
\end{lem}
\begin{proof}
    Let $\bM_0, \bM_1 \in \reals^{p \times p}$, then by reverse triangle inequality, we can obtain that,
    \begin{align*}
        \big\lvert \norm{\bM_0}_* - \norm{\bM_1}_* \big\rvert &\leq \norm{\bM_0 - \bM_1}_*
        \leq \sqrt{\rank\parent{\bM_0 - \bM_1}} \norm{\bM_0 - \bM_1}_F
        \leq p \norm{\bM_0 - \bM_1}_2,
    \end{align*}
    which proves the Lipschitz continuity of the nuclear norm operator with respect to the spectral norm of the input matrices.
\end{proof}
In addition to the above lemmas, it is straightforward that the sub-matrix operations utilized in Eq. \ref{eq:corr-rewrite} are Lipschitz continuous as they only extract the non-zero blocks in the target matrix. At this moment, we have shown that every component showing up in Eq. \ref{eq:corr-rewrite} is Lipschitz with respect to its inputs. Using the composition rule of Lipschitz functions, we have that the following three functions are Lipschitz to their inputs $\bZ$:
\begin{equation}
    t_1'(\bZ) = t_1(\bZ)_{[1:n, 1:n]}^{-1/2}, t_2'(\bZ) = t_2(\bZ)_{[n+1:2n, n+1:2n]}^{-1/2},t_3'(\bZ)=t_3(\bZ)_{[1:n, n+1:2n]},
\end{equation}
where the Lipschitz constants can be computed by chaining and multiplying the Lipschitz constants of each individual functions together and are given by $\sigma_{max}/\sigma_{min}^3$, $2\sigma_{max}$, and $\sigma_{max}/\sigma_{min}^3$, respectively.
\begin{remark}
One can easily verify that the function $t_1'(\cdot)$, $t_2'(\cdot)$, and $t_3'(\cdot)$ defined in the assumed domain are bounded using the Boundedness Theorem in mathematical analysis \citep{rudin1976principles}. This can be applied as all aforementioned functions are shown to be continuous and the domain of interests is assumed to be compact. For simplicity, we employ $M_1$, $M_2$, and $M_3$ as the upper bounds for their spectral norms in this paper.
\end{remark}
Furthermore, combining the product rules of Lipschitz functions together , it is straightforward that the canonical correlation estimation function $H(\cdot)_{n,p}$ defined on its domain of interest is Lipschitz continuous. By chaining all factors together and the product rule, the Lipschitz constant for $H(\cdot)$ can be expressed as follows:
\begin{equation}
    L = M_2(M_1 + M_3)\frac{\sigma_{max}}{\sigma_{min}^3}p + 2M_1M_3\sigma_{max}p.
\end{equation}
In the rest of this paper, we will simply utilize $L$ to denote the Lipschitz constant.

\subsection{Proof of Theorem 1}
\begin{thm} (Theorem 1 revisited)
    Let $\bff_{\hat\btheta} \in \cF$ denote the empirical risk minimizer. Then under certain regularity conditions, for any $\delta > 0$, the following bound holds with probability at least $1 - \delta$,
    \begin{align}
        \bbE[\cL(\bff_{\hat\btheta})] & - \inf_{\bff_{\btheta} \in \cF}\bbE[\cL(\bff_{\btheta})]
         \leq 4L \cR_{N_0}(\cF)+ \sqrt{\frac{2p^2\log(1/\delta)}{N_0}},\vspace{-0.04in}
    \end{align}
    where $p$ is the total correlation dimension, $N_0$ denote the size of dataset $\cD_0$, $L > 0$ is the Lipschitz constant of $H_{n,p}(\cdot)$, and $\cR_{N}(\cF)$ is the Rademacher Complexity of function class $\cF$ \citep{bartlett2002rademacher}. 
    \end{thm}
    \begin{remark}
    Instead of directly proving Theorem 1, we analyze the proposed extended loss function $\widetilde{L}(\bff, \bg)$ in terms of its generalization error and subsequently derive Theorem 1. This approach also yields results that are useful for proving Theorem 4.
    \end{remark}
    Let $\cD_0=\{(\bX_i, \bY_i) \sim \cP_{0}^{\otimes n}\}_{i=1}^{N_0}$ and $\cD_1=\{(\bX_i, \bY_i) \sim \cP_1^{\otimes n}\}_{i=1}^{N_1}$ denote the training dataset, where we slightly abuse the notation and let $\cP_1 = \cP_{1,x} \times \cP_{0,y}$ as introduced in Section 3. For simplicity purpose, we utilize $\bff_{\btheta}$ to collect parameters together and denote the combined function of $\bff$ and $\bg$, parameterized by $\btheta$. To revisit, the loss function with regularization is provided by:
    \begin{align}
         \wt\cL(\bff_{\btheta}) =& \ -\frac{1}{N_0}\sum_{(\bX, \bY) \in D_0} H_{n,p} \parent{\bff_{\btheta}(\bX,\bY)} +  \beta\frac{1}{N_1}\sum_{(\bX^*, \bY^*) \in D_1}H_{n,p} \parent{\bff_{\btheta}(\bX^*,\bY^*)}
    \end{align}
    Now, we define the empirical loss on the datasets $\cD_0$ and $\cD_1$ and the expected loss as follows, which are also known as risks in the context of empirical risk minimization:
    \begin{align}
        \hat R_{D_0, D_1}(\bff_{\btheta}) &= \wt\cL(\bff_{\btheta}) \text{ and }
        R(\bff_{\btheta}) = \bbE_{(\bX, \bY) \sim \cP_0^{\otimes n}, (\bX^*, \bY^*) \sim \cP_1^{\otimes n}}[\wt\cL(\bff_{\btheta})].
    \end{align}
    If we further decompose the two terms of the above loss function into two parts, we obtain,
    \begin{align}
        \hat R_{\cD_0, \cD_1}(\bff_{\btheta}) &= \hat R_{0,\cD_0}(\bff_{\btheta}) - \beta \hat R_{1,\cD_1}(\bff_{\btheta}) \text{ and }
        R(\bff_{\btheta}) = R_0(\bff_{\btheta}) - \beta R_1(\bff_{\btheta}),
    \end{align}
    where the risks for two terms are treated separately. To quantify the generalization errors for this extended objective function, we first provide the following lemmas.
    \begin{lem}\label{lem: risk-rewrite}
        Let $\bff_{\hat\btheta}, \bff_{\btheta^*}$ denote the minimizer to the empirical risk $\hat R_{\cD_0, \cD_1}(\bff_{\btheta})$ and expected risk $R(\bff_{\btheta})$, respectively. Then, the following inequality holds,
        \begin{align}
            R(\bff_{\hat\btheta}) \leq R(\bff_{\btheta^*}) + 2 \sup_{\bff_{\btheta}\in \cF}\Big  \lvert R_0(\bff_{\btheta}) - \hat R_{0, \cD_0}(\bff_{\btheta})\Big\rvert + 2\beta \sup_{\bff_{\btheta}\in \cF}\Big  \lvert R_1(\bff_{\btheta}) - \hat R_{1, \cD_1}(\bff_{\btheta})\Big\rvert,
        \end{align}
        which quantifies the optimality gap between $\bff_{\hat\btheta}$ and $\bff_{\btheta^*}$.
    \end{lem}
    \begin{proof}
    The lemma can be proved by simple algebraic manipulation and utilizing the fact that $\bff_{\btheta}$ is the empirical risk minimizer. Thus, we can obtain that,
        \begin{align*}
            R(\bff_{\hat\btheta}) - R(\bff_{\btheta^*}) &=  R(\bff_{\hat\btheta}) - \hat R_{\cD_0, \cD_1}(\bff_{\hat\btheta}) + \hat R_{\cD_0, \cD_1}(\bff_{\hat\btheta}) - R(\bff_{\btheta^*})\\
            &\leq R(\bff_{\hat\btheta}) - \hat R_{\cD_0, \cD_1}(\bff_{\hat\btheta}) + \hat R_{\cD_0, \cD_1}(\bff_{\btheta^*}) - R(\bff_{\btheta^*})\\
            &\leq 2 \sup_{\bff_{\btheta} \in \cF} \left\lvert R(\bff_{\btheta}) - \hat R_{\cD_0, \cD_1}(\bff_{\btheta}) \right\rvert\\
            &= 2 \sup_{\bff_{\btheta} \in \cF} \left\lvert R_0(\bff_{\btheta}) - \beta R_1(\bff_{\btheta}) - \parent{\hat R_{0,\cD_0}(\bff_{\btheta}) - \beta \hat R_{1,\cD_1}(\bff_{\btheta})} \right\rvert\\
            &\leq 2 \sup_{\bff_{\btheta} \in \cF} \left\lvert R_0(\bff_{\btheta}) - \hat R_{0,\cD_0}(\bff_{\btheta})\right\rvert + 2 \beta \sup_{\bff_{\btheta} \in \cF} \left\lvert R_1(\bff_{\btheta}) - \hat R_{1,\cD_1}(\bff_{\btheta}) \right\rvert.
        \end{align*}
        Rewriting and moving terms of the above inequality completes the proof of this lemma. Note that when the target loss function is the original function $\cL(\cdot)$ without regularization term, the upper bound reduces to simply $2 \sup_{\bff_{\btheta} \in \cF} \left\lvert R_0(\bff_{\btheta}) - \hat R_{0,\cD_0}(\bff_{\btheta})\right\rvert$.
    \end{proof}
    To establish a uniform convergence bound for quantifying the generalization errors of the proposed loss function, we make use of the Rademacher Complexity theory \citep{bartlett2002rademacher}, whose definition is provided below.
    \begin{defn}
        \label{defn: rmc}(Rademacher Complexity) \citep{bartlett2002rademacher} Let $\cF$ denote a class of mappings from $\cP$ to $\reals$, and let $S=\{s_1, ..., s_N\}$ be a sample from $\cP$. The empirical Rademacher complexity of the hypothesis class $\cF$ with respect to $S$ is defined as:
        \begin{equation}
            \hat\cR_S(\cF) = \bbE_{\bsigma}\left[\sup_{f \in \cF}\frac{1}{N}\sum_{i=1}^N \sigma_i f(s_i)\right]
        \end{equation}
        where $\bsigma =(\sigma_1, ...,\sigma_N)$ and $\{\sigma_i\}_{i=1}^N$ are independent Rademacher random variables that take values in $\{+1, -1\}$ uniformly. The Rademacher complexity is defined as:
        \begin{equation}
            \cR_N(\cF) = \bbE_{S \sim \cP^{\otimes N}}\left[\hat\cR_S(\cF)\right].
        \end{equation}
    \end{defn}
    Using the above definition, we can provide the following bound on $R_0(\bff_{\btheta})$ and $\hat R_{0, \cD_0}(\bff_{\btheta})$.

    \begin{thm}\label{thm: rade-norm}       For any $\delta > 0$, the following inequality holds for all $\bff_{\btheta}  \in \cF$ with probability at least $1 - \delta$,
        \begin{equation}
            R_0(\bff_{\btheta}) \leq \hat R_{0, \cD_0}(\bff_{\btheta}) + 2L\cR_{N_0}(\cF) + \sqrt{\frac{p^2\log(1/\delta)}{2N_0}}.
        \end{equation}
    \end{thm}
    \begin{proof}
    The proof of this theorem follows the standard procedure of proving uniform convergence bound which leverages McDiarmid's Inequality \citep{mcdiarmid1989inequality}. First, we define $\Phi(\cD_0) = \sup_{\bff_{\btheta}\in \cF} R_0(\bff_{\btheta}) - \hat R_{0, \cD_0}(\bff_{\btheta})$. Without losing generality, we define a dataset $\cD_0'$ to be a dataset that only differs from $\cD_0$ at the $i$th sample, that is, $\cD_0'=\parent{\cD_0 \setminus \{(\bX_i, \bY_i)\}}\cup \{(\bX^{'}, \bY^{'})\}$. Following this definition, we can obtain that,
        \begin{align*}
            \Phi(\cD_0) - \Phi(\cD_0') & \leq \sup_{\bff_{\btheta}\in \cF} \parent{R_0(\bff_{\btheta}) - \hat R_{0, \cD_0}(\bff_{\btheta})} - \sup_{\bff_{\btheta}\in \cF} \parent{R_0(\bff_{\btheta}) - \hat R_{0, \cD_0'}(\bff_{\btheta})}\\
            &\leq \sup_{\bff_{\btheta}\in \cF} \hat R_{0, \cD_0'}(\bff_{\btheta}) - \hat R_{0, \cD_0}(\bff_{\btheta})\\
            &= \sup_{\bff_{\btheta} \in \cF} -\frac{1}{N_0} \sum_{(\bX, \bY) \in D_0} H_{n,p} \parent{\bff_{\btheta}(\bX,\bY)} +\frac{1}{N_0} \sum_{(\bX, \bY) \in D_0'} H_{n,p} \parent{\bff_{\btheta}(\bX,\bY)}\\
            &= \sup_{\bff_{\btheta} \in \cF}\frac{1}{N_0}\parent{H_{n,p}(\bff_{\btheta}(\bX^{'}, \bY^{'})) - H_{n,p}\parent{\bff_{\btheta}(\bX_i,\bY_i)}}\\
            &\leq \frac{p}{N_0}.
        \end{align*}
        Utilizing the same structure, we obtain that $\Phi(\cD_0') - \Phi(\cD_0) \leq \frac{p}{N_0}$ and hence it is clear that $\Big\lvert \Phi(\cD_0) - \Phi(\cD_0')\Big\rvert \leq \frac{p}{N_0}$. Therefore, using McDiarmid's Inequality \citep{mcdiarmid1989inequality}, we have that for any $\delta > 0$, with probability at least $1 - \delta$,
        \begin{align}
            \Phi(\cD_0) \leq \bbE[\Phi(\cD_0)] + \sqrt{\frac{p^2 \log(1/\delta)}{2N_0}}.
        \end{align}
        Now, we define $\cD_0''=\{(\bX_i^{''}, \bY_i^{''})\}_{i=1}^{N_0}$ be another sequence of $N_0$ samples drawn from $\cP_{0}^{\otimes n}$ and denote the sample distribution with $\cP_\cD = \parent{\cP_0^{\otimes n}}^{\otimes N_0}$ for simplicity, then we have,\allowdisplaybreaks
        \begin{align*}
            \bbE[\Phi(\cD_0)] &= \bbE_{\cD_0 \sim \cP_\cD}\left[\sup_{\bff_{\btheta}\in \cF} R_0(\bff_{\btheta}) - \hat R_{0, \cD_0}(\bff_{\btheta})\right]\\
            &= \bbE_{\cD_0 \sim \cP_\cD}\left[\sup_{\bff_{\btheta}\in \cF} \bbE_{\cD_0'' \sim \cP_\cD} \left[\hat R_{0, \cD_0''}(\bff_{\btheta}) - \hat R_{0, \cD_0}(\bff_{\btheta})\right]\right]\\
            &\leq \bbE_{\cD_0 \sim \cP_\cD,\cD_0'' \sim \cP_\cD}\left[\sup_{\bff_{\btheta}\in \cF}  \hat R_{0, \cD_0''}(\bff_{\btheta}) - \hat R_{0, \cD_0}(\bff_{\btheta})\right]\\
            &=\bbE_{\cD_0 \sim \cP_\cD,\cD_0'' \sim \cP_\cD}\left[\sup_{\bff_{\btheta}\in \cF}  \frac{1}{N_0}\parent{\sum_{i=1}^{N_0}H_{n,p}(\bff_{\btheta}(\bX_i, \bY_i)) - H_{n,p}(\bff_{\btheta}(\bX_i^{''}, \bY_i^{''}))}\right]\\
            &= \bbE_{\cD_0 \sim \cP_\cD,\cD_0'' \sim \cP_\cD, \bsigma}\left[\sup_{\bff_{\btheta}\in \cF}  \frac{1}{N_0}\parent{\sum_{i=1}^{N_0}\sigma_i\parent{H_{n,p}(\bff_{\btheta}(\bX_i, \bY_i)) - H_{n,p}(\bff_{\btheta}(\bX_i^{''}, \bY_i^{''}))}}\right]\\
            &\leq  \bbE_{\cD_0 \sim \cP_\cD, \bsigma}\left[\sup_{\bff_{\btheta}\in \cF}  \frac{1}{N_0}\parent{\sum_{i=1}^{N_0}\sigma_iH_{n,p}(\bff_{\btheta}(\bX_i, \bY_i))}\right] \\
            &+ \bbE_{\cD_0'' \sim \cP_\cD, \bsigma}\left[\sup_{\bff_{\btheta}\in \cF}  \frac{1}{N_0}\parent{\sum_{i=1}^{N_0}-\sigma_iH_{n,p}(\bff_{\btheta}(\bX_i^{''}, \bY_i^{''}))}\right]\\
            &= \bbE_{\cD_0'' \sim \cP_\cD}\left[\hat\cR_{\cD_0''}(H_{n,p} \circ \cF)\right] + \bbE_{\cD_0 \sim \cP_\cD}\left[\hat\cR_{\cD_0}(H_{n,p} \circ \cF)\right]\\
            &= 2\cR_{N_0}(H_{n,p} \circ \cF)\\
            &= 2L \cR_{N_0}(\cF).
        \end{align*}
        Note that the above derivation utilizes standard results from the Rademacher Complexity theory \citep{bartlett2002rademacher} and elementary inequality such as Jensen's inequality. The last equality holds due to the Talagrand's contraction principle \citep{talagrand1996concentration}, where we have proved that $H_{n,p}(\cdot)$ is an $L$-Lipschitz function in Lemma \ref{lem:lipschitz}.
        Collecting the results together, we obtain that for any $\delta > 0$, with probability at least $1 - \delta$,
        \begin{equation}
            \Phi(\cD_0) \leq 2L \cR_{N_0}(\cF) + \sqrt{\frac{p^2 \log(1/\delta)}{2N_0}},
        \end{equation}
        and hence for all $\bff_{\btheta} \in \cF$, for any $\delta > 0$, with probability at least $1 - \delta$,
        \begin{align}
            R_0(\bff_{\btheta}) - \hat R_{0, \cD_0}(\bff_{\btheta}) \leq \Phi(\cD_0) \leq 2L\cR_{N_0}(\cF) + \sqrt{\frac{p^2\log(1/\delta)}{2N_0}},
        \end{align}
        which concludes the proof for this Theorem. Indeed, the same upper bound can also be derived for $\hat R_{0, \cD_0}(\bff_{\btheta}) - R_0(\bff_{\btheta})$ by simply changing the definition of $\Phi(\cD_0)$. These results will be essential when proving for Theorem 1.
    \end{proof}
    Chaining Lemma \ref{lem: risk-rewrite} and Theorem \ref{thm: rade-norm} together, the bound in Theorem 1 can be derived easily. Now, we keep analyzing the extended loss function in terms of its generalization errors to prepare us for the proof of Theorem 4. Similarly, we can derive a bound on $R_1(\bff_{\btheta})$ and $\hat R_{1, \cD_1}(\bff_{\btheta})$, for which the proof follows the exact same procedure as before.
    \begin{thm}\label{thm: rade-reg}
        For any $\delta > 0$, the following inequality holds for all $\bff_{\btheta}  \in \cF$ with probability at least $1 - \delta$,
        \begin{equation}
            R_1(\bff_{\btheta}) \leq \hat R_{1, \cD_1}(\bff_{\btheta}) + 2L\cR_{N_1}(\cF) + \sqrt{\frac{p^2\log(1/\delta)}{2N_1}}.
        \end{equation}
    \end{thm}
    The proof of the above Theorem A.3 follows the exact same structure of Theorem A.2 and is omitted here. Note that this Theorem A.3 will be utilized later together with Lemma \ref{lem: risk-rewrite}, Theorem \ref{thm: rade-norm}, Theorem 2, and Corollary 1 in the proof for Theorem 4.

\subsection{Proof of Theorem 2}
To prove Theorem 2, we first recall the well-known Matrix Bernstein inequality and derive its high probability form for later usage.
\begin{thm}\label{thm: matrix-bernstein}
    (Matrix Bernstein) \citep{tropp2015introduction} Let $\{\bS_i\}_{i=1}^n$ be a finite sequence of independent random matrices with common dimension $d_1 \times d_2$ where $\bbE[\bS_i]=0$ and $\norm{\bS_i} \leq L$ for all $i \in [n]$. Now, consider the random matrix $\bZ = \sum_{i=1}^n \bS_i$, then for all $t > 0$, it holds that,
    \begin{equation}
        \bbP(\norm{\bZ}_2 \geq t) \leq (d_1 + d_2) \exp\parent{\frac{-t^2/2}{\nu(\bZ) + Lt/3}},
    \end{equation}
    where $\nu(\bZ)$ denote the variance of matrix sum:
    \begin{align}
        \nu(\bZ) = \max \Big \{   \norm{\bbE \left[\bZ\bZ^\top\right]}_2, \norm{\bbE [\bZ^\top\bZ]}_2   \Big\}.
    \end{align}
\end{thm}

\begin{cor}\label{cor: matrix-bernstein-rewrite}
    (Matrix Bernstein in High-Probability Form) Under the same setting as in Theorem \ref{thm: matrix-bernstein}, for all $\delta > 0$, the following holds with probability at least $1 - \delta$:
    \begin{align}
        \norm{\bZ}_2 \leq \frac{2}{3}L \log\parent{\frac{d_1 + d_2}{\delta}} + \sqrt{2\nu(\bZ) \log\parent{\frac{d_1 + d_2}{\delta}}}
    \end{align}
\end{cor}

\begin{proof}
    The corollary can be proved by simply applying Theorem \ref{thm: matrix-bernstein} and finding an upper bound for the value of $t$ when the right-hand side is set to $\delta$. Therefore, we can obtain,
    \begin{align*}
        & (d_1 + d_2) \exp\parent{\frac{-t^2/2}{\nu(\bZ) + Lt/3}} = \delta\\
        \Longleftrightarrow \ \ & t^2 - \frac{2}{3}L\log\parent{\frac{d_1 + d_2}{\delta}}t - 2\log\parent{\frac{d_1 + d_2}{\delta}}\nu(\bZ) = 0
    \end{align*}
    Now, using the formula for roots of quadratic equation, we obtain that,
    \begin{equation}
        t = \frac{1}{2}\parent{\frac{2}{3}LA \pm \sqrt{\parent{\frac{2}{3}LA}^2 + 8 A\nu(\bZ)}},
    \end{equation}
    where we set $A = \log\parent{\frac{d_1 + d_2}{\delta}}$ for simplicity. In addition, we know that $A \geq 0$ as $d_1 + d_2 \geq 2$ and $\delta \in [0,1]$. Thus, it is clear that only the positive root survives and that,\allowdisplaybreaks
    \begin{align*}
        t &= \frac{1}{3}LA + \sqrt{\parent{\frac{1}{3}LA}^2 + 2 A\nu(\bZ)} \\
        &\leq \frac{1}{3}LA + \parent{\frac{1}{3}LA + \sqrt{2A\nu(\bZ)}}\\
        % &= \frac{2}{3}LA + \sqrt{2A\nu(\bZ)}\\
        &= \frac{2}{3}L\log\parent{\frac{d_1 + d_2}{\delta}} + \sqrt{2\nu(\bZ)\log\parent{\frac{d_1 + d_2}{\delta}}}.
    \end{align*}
    This completes the proof for the corollary.
\end{proof}
Now, applying this inequality, we can establish the non-asymptotic error bound for covariance estimate using a finite sample (in high-probability form), which is a standard result in literature \citep{tropp2015introduction} and hence proof is omitted.
\begin{lem}\label{lem: cov-est} (Non-asymptotic error bound for covariance estimate) \citep{tropp2015introduction}
    Let $\bx \in \reals^p$ be a centered random vector and $\bX\in \reals^{p \times n}$ be collections of their i.i.d. observations where $\norm{\bx_i}_2^2 \leq B$, and let $\hat\bSigma= \frac{1}{n}\bX\bX^\top$ be the $n$-sample finite estimate of covariance matrix of $\bx$. Then, for any $\delta > 0$, with probability at least $1 - \delta$,
    \begin{equation}
        \norm{\hat\bSigma - \bSigma}_2 \leq \frac{4B\log(2p / \delta)}{3n} + \sqrt{\frac{2B^2 \log (2p / \delta)}{n}}.
    \end{equation}
\end{lem}
Similarly, in this paper, we provide the following lemma for the non-asymptotic error bound for cross-covariance estimate using Corollary \ref{cor: matrix-bernstein-rewrite}.
\begin{lem}\label{lem: cross-cov-est}(Non-asymptotic error bound for cross-covariance estimate) 
    Let $\bx, \by \in \reals^p$ be two centered random vectors and $\bX, \bY \in \reals^{p \times n}$ be collections of their i.i.d. observations where $\norm{\bx_i}_2^2 \leq B_1, \norm{\by_i}_2^2 \leq B_2$, and let $\hat\bSigma_{12} = \frac{1}{n}\bX\bY^\top$ be the $n$-sample finite estimate of cross-covariance matrix of $\bx$ and $\by$. Then, for any $\delta > 0$, with probability at least $1 - \delta$,
    \begin{equation}
        \norm{\hat\bSigma_{12} - \bSigma_{12}}_2 \leq \frac{4\sqrt{B_1B_2}\log(2p / \delta)}{3n} + \sqrt{\frac{2B_1B_2 \log (2p / \delta)}{n}}.
    \end{equation}
\end{lem}
\begin{proof}
First, we define the following quantities:
    \begin{equation}
        \bS_i = \frac{1}{n}\parent{\bx_i\by_i^\top - \bSigma_{12}} \text{ and }\bZ = \sum_{i=1}^n \bS_i = \hat\bSigma_{12} - \bSigma_{12}.
    \end{equation}
    From the definition of $\bS_i$, we know that for all $i \in [n]$,
    \begin{align*}
        \bbE [\bS_i\bS_i^\top] &= \frac{1}{n^2} \bbE \left [ \parent{\bx_i\by_i^\top - \bSigma_{12}}\parent{\by_i\bx_i^\top - \bSigma_{12}^\top}\right]\\
      &= \frac{1}{n^2}\bbE \left[ \bx_i \parent{\by_i^\top\by_i}\bx_i^\top - \bSigma_{12}\by_i\bx_i^\top - \bx_i\by_i^\top\bSigma_{12}^\top + \bSigma_{12} \bSigma_{12}^\top\right]\\
      &= \frac{1}{n^2} \parent{\norm{\by_i}_2^2\bbE \left[ \bx_i\bx_i^\top  \right] - \Sigma_{12}\bbE\left[ \by_i\bx_i^\top\right] - \bbE\left[\bx_i\by_i^\top\right]\bSigma_{12}^\top + \bSigma_{12} \bSigma_{12}^\top}\\
      &= \frac{1}{n^2} \parent{\norm{\by_i}_2^2\bSigma_{11} - \Sigma_{12}\bSigma_{12}^\top- \bSigma_{12}\bSigma_{12}^\top + \bSigma_{12} \bSigma_{12}^\top}\\
      &= \frac{1}{n^2} \parent{\norm{\by_i}_2^2\bSigma_{11} - \Sigma_{12}\bSigma_{12}^\top}\\
      &\preceq \frac{\norm{\by_i}_2^2}{n^2} \bSigma_{11},
    \end{align*}
    where $\bA \preceq \bB$ means that the difference $\bB - \bA$ is positive semidefinite. The last step holds because $\frac{\norm{\by_i}_2^2}{n^2} \bSigma_{11} - \bbE [\bS_i\bS_i^\top] = \frac{1}{n^2}\bSigma_{12}\bSigma_{12}^\top$ can be identified as a Gram matrix which is always positive semidefinite. Therefore, we can obtain that for all $i \in [n]$, 
    \begin{equation}
        \norm{\bbE [\bS_i\bS_i^\top]}_2 \leq \frac{\norm{\by_i}_2^2}{n^2} \norm{\bSigma_{11}}_2 \leq \frac{B_1B_2}{n^2},
    \end{equation}
    where the last inequality holds by Jensen's inequality, that is,
    \begin{equation}
        \norm{\bSigma_{11}}_2 = \norm{\bbE \left[ \bx_i\bx_i^\top\right]}_2 \leq \bbE \left[ \norm{\bx_i\bx_i^\top}_2  \right] \leq B_1.
    \end{equation}
    A similar result can be derived for $\norm{\bbE [\bS_i^\top\bS_i]}_2$ for all $i \in [n]$. Now, utilizing the definition of matrix variance in Theorem \ref{thm: matrix-bernstein}, we can derive that,
    \begin{align*}
        \nu(\bZ) &= \max \Big \{   \norm{\bbE [\bZ\bZ^\top]}_2, \norm{\bbE [\bZ^\top\bZ]}_2   \Big\}\\
        &= \max \left \{   \norm{\bbE \left[\sum_{i=1}^n \bS_i\bS_i^\top\right]}_2, \norm{\bbE \left[\sum_{i=1}^n \bS_i^\top\bS_i \right]}_2   \right\}\\
        &\leq \frac{B_1B_2}{n},
    \end{align*}
    where the last inequality is true because by triangle inequality,
    \begin{align*}
        \norm{\bbE \left[\sum_{i=1}^n \bS_i\bS_i^\top\right]}_2 &= \norm{\sum_{i=1}^n \bbE \left[\bS_i\bS_i^\top\right]}_2
        \leq\sum_{i=1}^n  \norm{\bbE [\bS_i\bS_i^\top]}_2 \leq \frac{B_1B_2}{n},\\
        \norm{\bbE \left[\sum_{i=1}^n \bS_i^\top\bS_i\right]}_2 &= \norm{\sum_{i=1}^n \bbE \left[\bS_i^\top\bS_i\right]}_2
        \leq\sum_{i=1}^n  \norm{\bbE [\bS_i^\top\bS_i]}_2 \leq \frac{B_1B_2}{n}.
    \end{align*}
    Furthermore, using Jensen's inequality again, we obtain,
    \begin{align}
        \norm{\bSigma_{12}}_2 &= \norm{\bbE \left[ \bx_i\by_i^\top\right]}_2 \leq \bbE \left[ \norm{\bx_i\by_i^\top}_2  \right] \leq \sqrt{B_1B_2},
    \end{align}
    and hence for all $i \in [n]$, the 2-norm of $\bS_i$ is bounded by:
    \begin{equation}
        \norm{\bS_i}_2 = \frac{1}{n} \norm{\bx_i\by_i^\top - \bSigma_{12}}_2 \leq \frac{1}{n}\parent{\norm{\bx_i\by_i^\top}_2 + \norm{\bSigma_{12}}_2} \leq \frac{2 \sqrt{B_1B_2}}{n}.
    \end{equation}
    Finally, we can plug in everything we have derived into Corollary \ref{cor: matrix-bernstein-rewrite} and obtain that for any $\delta > 0$, it holds with probability at least $1 - \delta$ that,
    \begin{align}
        \norm{\hat\bSigma_{12} - \bSigma_{12}}_2 \leq \frac{4\sqrt{B_1B_2}\log(2p / \delta)}{3n} + \sqrt{\frac{2B_1B_2 \log (2p / \delta)}{n}}.
    \end{align}
\end{proof}

\begin{remark}
    Before deriving the non-asymptotic error bound for canonical correlation estimation, it is important to note that assuming the true covariance matrix is bounded in terms of its eigenvalues is reasonable. This assumption aligns naturally with practical scenarios that this paper deals with, where high-dimensional sensor data usually exhibit a pre-defined range. Similarly, we can assume the boundedness of eigenvalues of the estimated covariance matrix. Let $\bX \in \reals^{p \times n}$ denote the collected data matrix of $n$ samples and let $\sigma_{min} \leq \sigma(\bX) \leq \sigma_{max}$ denote the range of its singular values, then it is clear that for fixed feature dimension $p$, both $\sigma_{min}$ and $\sigma_{max}$ scales up in a rate of $O(\sqrt{n})$. Therefore, we can easily verify that the eigenvalues of the finite sample estimate $\frac{1}{n}\bX\bX^\top$ have an order of $O(1)$, rendering our assumption reasonable. Therefore, in this paper, we assume that $\lambda_{min} \leq \lambda(\bSigma_{11}), \lambda(\bSigma_{12}), \lambda(\bSigma_{12}), \lambda(\hat\bSigma_{11}), \lambda(\hat\bSigma_{22}), \lambda(\hat\bSigma_{12}) \leq \lambda_{max}$, where the definitions of those covariances matrix are introduced in Section 2. In addition, it is noteworthy that the singular values and eigenvalues coincide for those positive definite matrices so we will use the notations interchangeably later.
\end{remark}

\begin{thm} (Non-asymptotic error bound for canonical correlation estimate; Theorem 2 revisited)
    Let $\rho^*$ denote the true $p$-dimensional canonical correlation coefficient between two random vectors $\bx$ and $\by$. Let $\{(\bx_i, \by_i)\}_{i \in [n]}$ be an i.i.d. finite $n$-sample estimate with $\norm{\bx_i}_2^2, \norm{\by_i}_2^2\leq B$ and $\bX, \bY$ be the collected data matrix, then with probability at least $1 - \delta$, 
    \begin{equation}
        \Big\lvert H_{n,p}(\bX, \bY) - \rho^* \Big\rvert \leq \frac{4Cp\log(6p / \delta)}{3n} + \sqrt{\frac{2C^2p^2\log (6p / \delta)}{n}},
    \end{equation}
    where $C$ is a constant function of $B, \lambda_{min}$ and $\lambda_{max}$, where the latter two denote the range of eigenvalues of the covariance and cross-covariance matrix. 
\end{thm}
\begin{proof}
First, we rewrite the desired quantity using triangle inequality and some simple algebraic manipulations as follows:
    \begin{align*}
        & \bigg\lvert \norm{\hat\bSigma_{11}^{-1/2}\hat\bSigma_{12} \hat\bSigma_{22}^{-1/2}}_* -\norm{\bSigma_{11}^{-1/2}\bSigma_{12} \bSigma_{22}^{-1/2}}_*  \bigg\rvert \\
        \leq \ & \norm{\hat\bSigma_{11}^{-1/2}\hat\bSigma_{12} \hat\bSigma_{22}^{-1/2} -\bSigma_{11}^{-1/2}\bSigma_{12} \bSigma_{22}^{-1/2}}_* \\
        =\ &\norm{\parent{\hat\bSigma_{11}^{-1/2}-\bSigma_{11}^{-1/2}}\hat\bSigma_{12} \hat\bSigma_{22}^{-1/2}}_* \tag{T1}\\
        + \ & \norm{\bSigma_{11}^{-1/2}\parent{\hat\bSigma_{12}-\bSigma_{12}} \hat\bSigma_{22}^{-1/2}}_* \tag{T2}\\
        + \ & \norm{\bSigma_{11}^{-1/2}\bSigma_{12} \parent{\hat\bSigma_{22}^{-1/2} - \bSigma_{22}^{-1/2}}}_* \tag{T3}.
    \end{align*}
    Now, it suffices to investigate the error bound for those three terms (T1), (T2), and (T3), separately. For (T1), we obtain that,
    \begin{align*}   \norm{\parent{\hat\bSigma_{11}^{-1/2}-\bSigma_{11}^{-1/2}}\hat\bSigma_{12} \hat\bSigma_{22}^{-1/2}}_* &\leq p\norm{\parent{\hat\bSigma_{11}^{-1/2}-\bSigma_{11}^{-1/2}}\hat\bSigma_{12} \hat\bSigma_{22}^{-1/2}}_2\\
        & \leq p \norm{\hat\bSigma_{11}^{-1/2}-\bSigma_{11}^{-1/2}}_2\norm{\hat\bSigma_{12}}_2\norm{\hat\bSigma_{22}^{-1/2}}_2\\
        & \leq p \lambda_{max}\parent{\frac{1}{\sqrt{\lambda_{min}}}}\norm{\hat\bSigma_{11}^{-1/2}-\bSigma_{11}^{-1/2}}_2\\
        &\leq p \lambda_{max}\parent{\frac{1}{\sqrt{\lambda_{min}}}}\parent{\frac{1}{2\lambda_{min}^{3/2}}}\norm{\hat\bSigma_{11} - \bSigma_{11}}_2\\
        &= \frac{p\lambda_{max}}{2\lambda_{min}^2}\norm{\hat\bSigma_{11} - \bSigma_{11}}_2,
    \end{align*}
    where the second last inequality comes from Lemma \ref{lem: lipschitz-inv-sqrt}. Similarly, for (T2), we have that,
    \begin{align*}
\norm{\bSigma_{11}^{-1/2}\parent{\hat\bSigma_{12}-\bSigma_{12}} \hat\bSigma_{22}^{-1/2}}_* &\leq p\norm{\bSigma_{11}^{-1/2}\parent{\hat\bSigma_{12}-\bSigma_{12}} \hat\bSigma_{22}^{-1/2}}_2\\
        &\leq p\norm{\bSigma_{11}^{-1/2}}_2\norm{\hat\bSigma_{22}^{-1/2}}_2\norm{\hat\bSigma_{12}-\bSigma_{12}}_2 \\
        &\leq p \parent{\frac{1}{\sqrt{\lambda_{min}}}}\parent{\frac{1}{\sqrt{\lambda_{min}}}}\norm{\hat\bSigma_{12}-\bSigma_{12}}_2\\
        &\leq \frac{p}{\lambda_{min}}\norm{\hat\bSigma_{12}-\bSigma_{12}}_2,
    \end{align*}
    and for (T3), we can obtain,
    \begin{align*}
        \norm{\bSigma_{11}^{-1/2}\bSigma_{12} \parent{\hat\bSigma_{22}^{-1/2} - \bSigma_{22}^{-1/2}}}_* &\leq p\norm{\bSigma_{11}^{-1/2}\bSigma_{12} \parent{\hat\bSigma_{22}^{-1/2} - \bSigma_{22}^{-1/2}}}_2\\
        &\leq p\norm{\bSigma_{11}^{-1/2}}_2\norm{\bSigma_{12}}_2\norm{\hat\bSigma_{22}^{-1/2} - \bSigma_{22}^{-1/2}}_2\\
        &\leq p \parent{\frac{1}{\sqrt{\lambda_{min}}}}\lambda_{max}\norm{\hat\bSigma_{22}^{-1/2} - \bSigma_{22}^{-1/2}}_2\\
        &\leq p \parent{\frac{1}{\sqrt{\lambda_{min}}}}\lambda_{max}\parent{\frac{1}{2\lambda_{min}^{3/2}}}\norm{\hat\bSigma_{22} - \bSigma_{22}}_2\\
        &= \frac{p\lambda_{max}}{2\lambda_{min}^2}\norm{\hat\bSigma_{22} - \bSigma_{22}}_2.
    \end{align*}
    Now, we can apply Lemma \ref{lem: cov-est} and Lemma \ref{lem: cross-cov-est} together with union bound to derive the final error bound for (T1), (T2), and (T3), respectively. For instance, for any $\delta > 0$, the following holds for (T1) with probability at least $1 - \delta$,
    \begin{align}
        \norm{\parent{\hat\bSigma_{11}^{-1/2}-\bSigma_{11}^{-1/2}}\hat\bSigma_{12} \hat\bSigma_{22}^{-1/2}}_* &\leq \frac{p\lambda_{max}}{2\lambda_{min}^2}\parent{\frac{4B\log(2p / \delta)}{3n} + \sqrt{\frac{2B^2 \log (2p / \delta)}{n}}}.
    \end{align}
    Similar bounds can be derived for the other two terms in the same way. Thus, combining all results together and by union bounds, we can have that for all $\delta > 0$, the following holds with probability at least $1 - \delta$:
    \begin{align}
        & \bigg\lvert \norm{\hat\bSigma_{11}^{-1/2}\hat\bSigma_{12} \hat\bSigma_{22}^{-1/2}}_* -\norm{\bSigma_{11}^{-1/2}\bSigma_{12} \bSigma_{22}^{-1/2}}_*  \bigg\rvert \\
        &\leq \parent{\frac{p\lambda_{max}}{\lambda_{min}^2} + \frac{p}{\lambda_{min}}}\parent{\frac{4B\log(6p / \delta)}{3n} + \sqrt{\frac{2B^2 \log (6p / \delta)}{n}}}\\
        &= \frac{4Cp\log(6p / \delta)}{3n} + \sqrt{\frac{2C^2p^2\log (6p / \delta)}{n}},
    \end{align}
    where we utilize $\delta_1=\delta_2=\delta_3 = \delta/3$ when computing the error bound for each term and the constant is given as follows:
    \begin{equation}
        C = \frac{B(\lambda_{min} + \lambda_{max})}{\lambda_{min}^2}.
    \end{equation}
    This concludes the proof of this non-asymptotic error bound.
\end{proof}
\begin{remark}
    It is worth mentioning that this error bound has an order of $O(\frac{1}{\sqrt{n}})$, which can be utilized to analyze its asymptotic behavior later.
\end{remark}

\subsection{Proof of Corollary 1}\label{apd: cor-1-proof}
Inherited the notations in Section 3, for any given $\bff_{\btheta} \in \cF$, let $\wt\cP_{0}(\bff_{\btheta})$ be the corresponding feature distribution of normal process signature signals and normal quality data ($\cP_{0}$) induced by the feature extraction function $\bff_{\btheta}$, and for $(\bu, \bv) \sim \wt\cP_{0}(\bff_{\btheta})$, let $\rho^*_{0,\bff_{\btheta}}$ denote the true canonical correlation coefficients between random vectors $(\bu, \bv)$ (resp.  $\wt\cP_{1}(\bff_{\btheta})$ and $\rho^*_{1,\bff_{\btheta}}$). Furthermore, for all $\delta > 0$, we employ $\xi_0(\delta;\bff_{\btheta})$ and $\xi_1(\delta;\bff_{\btheta})$ to denote the non-asymptotic error between the $n$-sample estimates and true correlations for $(\bu, \bv) \sim \wt\cP_{0}(\bff_{\btheta})$ and $(\bu, \bv) \sim \wt\cP_{1}(\bff_{\btheta})$, respectively.
\begin{cor}(Corollary 1 revisited) Under certain regularity conditions, for all $\delta > 0$, the following bound holds with probability at least $1 - \delta$,\allowdisplaybreaks
    \begin{equation}
        \Big\lvert \inf_{\bff_{\btheta} \in \cF}\bbE[\cL(\bff_{\btheta})] - \inf_{\bff_{\btheta} \in \cF} \bbE[\cL^*(\bff_{\btheta})]\Big\rvert \leq \xi_0^*(\delta),\vspace{-0.05in}
    \end{equation}
    where $\xi_0^*({\delta}) \sim \cO(n^{-1/2})$ is a function of $\delta$ constructed via the non-asymptotic error bounds $\xi_0(\delta;\bff_{\btheta})$ for $\bff_{\btheta} \in \cF$ and $\epsilon$-net techniques \citep{haussler1992decision}.
\end{cor}

To prove Corollary 1, we first provide the following lemma to quantify the bound between $\bbE\left[\wt\cL(\bff_{\btheta})\right]$ and $ \bbE\left[\wt\cL^*(\bff_{\btheta})\right]$ using Theorem 2. Note that here we use the same strategy when proving for Theorem 1, where we first focus on the extended loss function $\widetilde\cL(\bff_{\btheta})$ and subsequently derive Corollary 1. 
\begin{lem}\label{lem: nonasymptotic}
    For all $\bff_{\btheta} \in \cF$, for any $\delta > 0$, it holds with probability at least $1 - \delta$,
    \begin{equation*}
        \Big\lvert \bbE\left[\wt\cL(\bff_{\btheta})\right] - \bbE\left[\wt\cL^*(\bff_{\btheta})\right]\Big\rvert \leq \xi(\delta;\bff_{\btheta}),
    \end{equation*}
    where $\xi(\delta;\bff_{\btheta}) = \xi_0(\delta/2;\bff_{\btheta}) + \beta\xi_1(\delta/2;\bff_{\btheta})$ denote the combined error.
\end{lem}
\begin{proof}\allowdisplaybreaks
The proof utilizes Jensen's inequality and results from Theorem 2
    \begin{align*}
        \Big\lvert \bbE\left[\wt\cL(\bff_{\btheta})\right] -\bbE\left[\wt\cL^*(\bff_{\btheta})\right]\Big\rvert &= \Big\lvert \bbE\left[\wt\cL(\bff_{\btheta}) - \wt\cL^*(\bff_{\btheta})\right]\Big\rvert\\
        &\leq  \bbE\left[\Big\lvert\wt\cL(\bff_{\btheta}) - \wt\cL^*(\bff_{\btheta})\Big\rvert\right]\\
        &= \bbE\left[\Big\lvert\hat R_{0,\cD_0}(\bff_{\btheta}) -\beta \hat R_{1,\cD_1}(\bff_{\btheta}) +\rho^*_{0, \bff_{\btheta}} - \beta\rho^*_{1, \bff_{\btheta}}\Big\rvert\right]\\
        &\leq \bbE\left[\Big\lvert\hat R_{0,\cD_0}(\bff_{\btheta}) +\rho^*_{0, \bff_{\btheta}}\Big\rvert  +
         \beta\Big\lvert-\hat R_{1,\cD_1}(\bff_{\btheta}) - \rho^*_{1, \bff_{\btheta}}\Big\rvert\right]\\
         &= \bbE\left[\frac{1}{N_0}\Bigg\lvert\sum_{(\bX, \bY)  \in \cD_0}\parent{\rho^*_{0, \bff_{\btheta}} - H_{n,p}(\bff_{\btheta}(\bX, \bY))}\Bigg\rvert\right]\\
         &+ \beta\bbE\left[\frac{1}{N_1}\Bigg\lvert\sum_{(\bX^*, \bY^*)  \in \cD_1}\parent{H_{n,p}(\bff_{\btheta}(\bX^*, \bY^*)) - \rho^*_{1, \bff_{\btheta}}}\Bigg\rvert\right]\\
         &\leq \bbE\left[\frac{1}{N_0}\sum_{(\bX, \bY)  \in \cD_0}\Big\lvert H_{n,p}(\bff_{\btheta}(\bX, \bY)) - \rho^*_{0, \bff_{\btheta}} \Big\rvert\right]\\
         &+ \beta\bbE\left[\frac{1}{N_1}\sum_{(\bX^*, \bY^*)  \in \cD_1}\Big\lvert H_{n,p}(\bff_{\btheta}(\bX^*, \bY^*)) - \rho^*_{1, \bff_{\btheta}}\Big\rvert\right]\\
         &\leq \xi_0(\delta/2;\bff_{\btheta}) + \beta\xi_1(\delta/2;\bff_{\btheta})\\
         &= \xi(\delta;\bff_{\btheta}),
    \end{align*}
    where the second last inequality holds with probability at least $1 - \delta$ by union bound.
\end{proof}
\allowdisplaybreaks
With this Lemma \ref{lem: nonasymptotic}, we can quantify the gap between the expected loss and the theoretical loss at optimality for the extended objective function $\widetilde\cL(\bff_{\btheta})$. As mentioned earlier, these results will be used to prove for Corollary 1. Let $S = \left\{\parent{\bbE[\wt\cL(\bff_{\btheta})], \bbE[\wt\cL^*(\bff_{\btheta})]}: \bff_{\btheta} \in \cF\right\}$ denote the paired set of the expected loss calculated using $n$-sample estimates and its corresponding theoretical loss. We further assume that the set $S$ is compact, which indicates that we can find a finite subcover of the set. 
\begin{remark}
Note that this compactness assumption is quite natural, as the elements of the set $S$ are closely related to the extracted features from sensing data collected in real-world manufacturing processes, which usually exhibit a finite range.    
\end{remark} 
Now, we construct a standard $\epsilon$-net \citep{haussler1992decision} $\cX$ on the target set $S$ such that for any $s \in S$, there exist $s' \in \cX$ such that $s \in B(s', \epsilon)$, where $B(s', \epsilon)$ denotes the $\epsilon$-ball centered at $s'$. In plain words, this means that for any element $s \in S$, we can find an element in $\cX$ such that $s$ is inside its neighborhood. Without losing of generality, we denote these representative points in the $\epsilon$-net as follows:
    \begin{align*}
        \cX = \left\{\parent{\bbE[\wt\cL(\bff_{\btheta_i})], \bbE[\wt\cL^*(\bff_{\btheta_i})]}: i \in [N_\epsilon]\right\},
    \end{align*}
    where $N_\epsilon$ is the number of points in the constructed $\epsilon$-net. In addition, we collect the corresponding $\bff_{\btheta_i}$ of those representatives into a new set $\cF_{\epsilon} = \{\bff_{\btheta_i}: i \in [N_\epsilon]\}$. Applying Lemma \ref{lem: nonasymptotic} and union bound on those representative points, we can obtain that for any $\delta > 0$, we have with probability at least $1 - \delta$ that,
    \begin{align*}
        \Big\lvert \bbE[\wt\cL(\bff_{\btheta})] - \bbE[\wt\cL^*(\bff_{\btheta})]\Big\rvert  \leq \xi(\delta / N_{\epsilon}; \bff_{\btheta}) , \forall \bff_{\btheta} \in \cF_\epsilon.
    \end{align*}
    Now, we investigate the relationship between representative points (i.e. $\cX$) and all other points in the target set $S$. For simplicity, we utilize $s=(a,b)$ to denote an element in $S$. Similarly, let $s'=(a',b') \in \cX$ denotes its closest representative points in the $\epsilon$-net, then by triangle inequality, we obtain that,
    \begin{align*}
        \lvert a - b\rvert &= \lvert (a - a') + (a' - b') + (b' - b)\rvert \leq \lvert a- a' \rvert + \lvert a'- b' \rvert + \lvert b- b' \rvert \leq \lvert a'- b' \rvert + 2\epsilon,
    \end{align*}
    because $\sqrt{(a - a')^2 + (b - b')^2} \leq \epsilon$ according to the design of $\cX$. Therefore, combining these results together, we obtain that for all $\bff_{\btheta} \in \cF$, for all $\delta > 0$, it holds with probability at least $1 - \delta$ that,
    \begin{align*}
        \sup_{\bff_{\btheta}  \in \cF}\Big\lvert \bbE[\wt\cL(\bff_{\btheta})] - \bbE[\wt\cL^*(\bff_{\btheta})]\Big\rvert &\leq \sup_{\bff_{\btheta}  \in \cF_\epsilon}\Big\lvert \bbE[\wt\cL(\bff_{\btheta})] - \bbE[\wt\cL^*(\bff_{\btheta})]\Big\rvert + 2\epsilon
        \leq \sup_{\bff_{\btheta}  \in \cF_\epsilon}\xi(\delta / N_{\epsilon}; \bff_{\btheta}) + 2\epsilon.
    \end{align*}
    Then, utilizing elementary knowledge in mathematical analysis, we can have that for any $\delta > 0$, the following bound holds with probability at least $1 - \delta$,
    \begin{align*}
         \Big\lvert \inf_{\bff_{\btheta} \in \cF}\bbE[\wt\cL(\bff_{\btheta})] - \inf_{\bff_{\btheta} \in \cF} \bbE[\wt\cL^*(\bff_{\btheta})]\Big\rvert &\leq \sup_{\bff_{\btheta}  \in \cF}\Big\lvert \bbE[\wt\cL(\bff_{\btheta})] - \bbE[\wt\cL^*(\bff_{\btheta})]\Big\rvert
         \leq \xi^*(\delta),
    \end{align*}
    where we set $\xi^*(\delta) = \sup_{\bff_{\btheta}  \in \cF_\epsilon}\xi(\delta / N_{\epsilon}; \bff_{\btheta}) + 2\epsilon$ for simplicity. Note that here $\xi^*(\delta)$ shares the same order as the non-asymptotic error bound mentioned in Theorem 2. At this stage, we finished the analysis for the extended function. To prove for Corollary 1, it just follows the same procedure but the upper bound becomes $\xi_0^*(\delta) = \sup_{\bff_{\btheta}  \in \cF_\epsilon}\xi_0(\delta / N_{\epsilon}; \bff_{\btheta}) + 2\epsilon$, where the corresponding modified $\epsilon$-net can be simply defined as $\cX_0 = \left\{\parent{\bbE[\cL(\bff_{\btheta_i})], \bbE[\cL^*(\bff_{\btheta_i})]}: i \in [N_\epsilon]\right\}$.
    
\subsection{Proof of Theorem 3}
\begin{thm}
Under the same setting described in Theorem 1 and Corollary 1, for any $\delta > 0$, the following holds with probability at least $1 - \delta$,
\begin{align}
     \bbE[\cL(\bff_{\hat\btheta})] \leq \inf_{\bff_{\btheta} \in \cF} &\bbE[\cL^*(\bff_{\btheta})] + \xi_0^{*}(\delta/2) + 4L \cR_{N_0}(\cF)+ \sqrt{\frac{2p^2\log(2/\delta)}{N_0}},
\end{align}
which provides an upper bound for the proposed loss function in the empirical setting. Furthermore, as $\xi_0^*(\delta/2) \sim \cO(n^{-1/2})$, it holds that asymptotically, the error terms will vanish as the number of training samples ($N_0$) and the number of samples within each window ($n$) used to estimate canonical correlations are sufficiently large.
\end{thm}

\begin{proof}
The proof of this theorem just requires combining the results from Theorem 1 and Corollary 1 with union bounds and is omitted due to its triviality.
\end{proof}

\subsection{Proof of Theorem 4}
\begin{assum}
(Key assumption revisited; Existence of separable mapping) There exists $\bff_{\btheta^\circ} \in \cF$ such that $\rho^*_{0,\bff_{\btheta^\circ}} - \rho^*_{1,\bff_{\btheta^\circ}} \geq cp$, where $c \in [0,1]$ quantifies the gap of canonical correlations between data coming from the normal ($\cP_0$) and abnormal ($\cP_1$) distribution.
\end{assum}
\begin{cor}\label{cor: cp}
Under the data separability assumption, the following always hold,
    \begin{align*}
        \inf_{\bff_{\btheta} \in \cF} \bbE[\wt\cL^*(\bff_{\btheta})] \leq -cp.
    \end{align*}
\end{cor}
\begin{proof}
The proof utilizes the property of infimum and is given as follows:
    \begin{align}
         \inf_{\bff_{\btheta} \in \cF} \bbE[\wt\cL^*(\bff_{\btheta})] \leq \bbE[\wt\cL^*(\bff_{\btheta^\circ})] = -\rho^*_{0,\bff_{\btheta^\circ}} +\beta \rho^*_{1,\bff_{\btheta^\circ}}\leq -\rho^*_{0,\bff_{\btheta^\circ}} + \rho^*_{1,\bff_{\btheta^\circ}}\leq -cp,
    \end{align}
    where the second last step uses the fact that $\beta \in [0,1]$. 
\end{proof}

\begin{thm} (Theorem 4 revisited)
Under the same setting as Theorem \ref{thm: rmc} but with the extended objective function, for any $\delta > 0$, the following holds with probability at least $1 - \delta$,
\begin{align}
     \bbE[\wt\cL(\bff_{\hat\btheta})] \leq -cp &+ \xi^{*}(\delta/2) + 4L\parent{\cR_{N_0}(\cF)+\beta\cR_{N_1}(\cF)}\notag\\
     &+ \sqrt{2\log(2/\delta)}p\parent{N_0^{-1/2} + \beta N_1^{-1/2}}, 
\end{align}
where $\xi^*(\delta / 2)$ is a function determined by the non-asymptotic errors derived in Theorem 2 and its complete form is provided in Appendix A.3. This Theorem provides an upper bound for the expectation of the modified objective with regularization terms. After decomposition, this states that the score gap between normal and abnormal distributions is preserved under the empirical minimizer $\bff_{\hat\btheta}$.
\end{thm} 

\begin{proof}
    The proof of this theorem requires chaining the results from Theorem 1, Corollary 1, and Corollary 2. First, we obtain the following by combining the two corollaries together.
    \begin{align*}
        \inf_{\bff_{\btheta} \in \cF}\bbE[\wt\cL(\bff_{\btheta})] &\leq \inf_{\bff_{\btheta} \in \cF} \bbE[\wt\cL^*(\bff_{\btheta})] + \xi_0^*(\delta/2) \leq -cp + \xi_0^*(\delta/2).
    \end{align*}
    Now, using the generalization bound (Theorem 1) derived before, we have that for any $\delta > 0$, the following holds with probability at least $1 - \delta$:
    \begin{align*}
        \bbE[\wt\cL(\bff_{\hat\btheta})] &\leq \inf_{\bff_{\btheta} \in \cF}\bbE[\wt\cL(\bff_{\btheta})] + 4L \parent{\cR_{N_0}(\cF)+\beta\cR_{N_1}(\cF)} + \sqrt{2\log(2/\delta)}p\parent{\sqrt{\frac{1}{N_0}} + \beta\sqrt{\frac{1}{N_1}}},
    \end{align*}
    where union bounds are applied to combine previous results. This concludes the proof of this Theorem.
\end{proof}
\begin{remark}
    Now, it is easy to observe that the expectation of extended objective function under the empirical minimizer $\bff_{\hat\btheta}$ can be written as follows:
    \begin{align*}
        \bbE[\wt\cL(\bff_{\hat\btheta})] = -\bbE_{(\bX, \bY) \sim \cP_0^{\otimes n}}\left[H_{n,p} \parent{\bff_{\hat\btheta}(\bX,\bY)}\right] +  \beta\bbE_{(\bX^*, \bY^*) \sim \cP_1^{\otimes n}}\left[H_{n,p} \parent{\bff_{\hat\btheta}(\bX^*,\bY^*)}\right],
    \end{align*}
    which is essentially the expectation of loss under the empirical risk minimizer $\bff_{\hat\btheta}$. Therefore, in plain words, an upper bound on it indicates that in the testing time, there will exist a gap between correlation scores of normal and abnormal data.
\end{remark}

%% file: src/appendix/simulation.tex
\newpage
\section{Simulation}

\subsection{Data Generation Process} \label{apd: simulatin-dgp}
The data generation process of this simulation study is shown in Figure \ref{fig: apd-simulation-dgp}. Table \ref{apd: tab-heuristics} presents the detailed heuristic functions used for transforming process signature features to the actual signals and generating quality images. The autoencoder architectures employed in this data generation process uses convolutional neural networks as backbones \citep{lecun2015deep} and are described below:
\begin{figure}[t]
    \centering
    \includegraphics[width=1.0\linewidth]{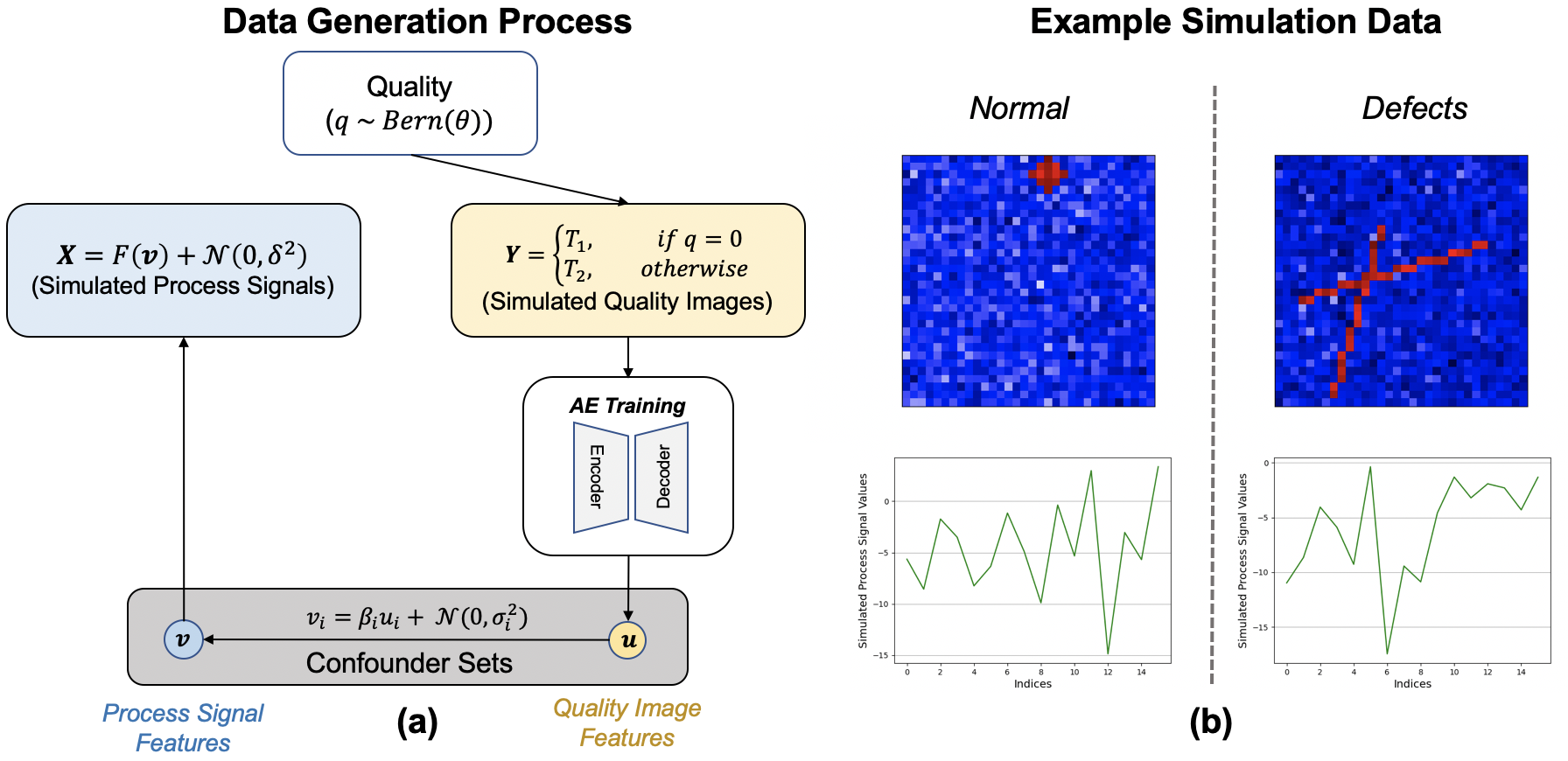}
    \caption{(Figure 5 revisited) Illustration of the data generation process used in the simulation experiments
(left) and a visualization of example simulation data (right).}
    \label{fig: apd-simulation-dgp}
\end{figure}
\begin{itemize}
    \item \textbf{Encoder:} The encoder network consists of two convolutional layers (\texttt{Conv2d}) with kernel size 5, each followed by a max-pooling layer (\texttt{MaxPool2d}). The convolutional layers project the input from a single channel to 32 and subsequently to 64 feature maps, with each pooling operation reducing the spatial dimensions by a factor of two. The resulting feature maps are then flattened and passed through two fully connected layers with ReLU activation in between, producing outputs of dimension 512 and a configurable hidden feature dimension, respectively.
    \item \textbf{Decoder:} The decoder structure mirroring that of the aforementioned encoder model. It begins with two fully connected layers: the first maps the input from a customizable hidden feature dimension to 512 units, and the second expands this to match the flattened spatial dimensions of the encoded feature maps (i.e., 64 × 5 × 5). The output is then reshaped and passed through two transposed convolutional layers (\texttt{ConvTranspose2d}), which progressively reconstruct the spatial dimensions of the original image. Between these layers, max-unpooling operations (\texttt{MaxUnpool2d}) are applied using the stored pooling indices from the encoder to restore the original resolution. A final sigmoid activation is applied to constrain the output values to the original pixel range of [0, 1].
\end{itemize}
In addition, in this simulation study, the hidden feature dimension is selected to be $ p = 6$ with the linear mapping coefficient $\bbeta = [0.3, 1, 2, 0.4, -0.5, -0.7]^\top$ and noise level $\sigma_i = 0.1$. The parameters of generation functions in Table \ref{apd: tab-heuristics} are set to $n = 3$, $d_0=32$, $lb=2$, $ub=6$, $t=2$, $h=6$, $h_1=32$, and $h_2=16$. In this simulation, when generating simulated process signature signals from the underlying features $\bv$, the noise level $\delta$ is gradually varied to evaluate the robustness of the proposed method to noise and its capability to analyze signals of varying complexity.

\begin{table}[t!]
\centering\small
\caption{Heuristic functions used for generating simulation data. When generating images, a $d_0 \times d_0$ blank image is provided as the background before any shapes are generated.}
\begin{tabular}{|>{\centering\arraybackslash}m{2.5cm}|>{\centering\arraybackslash}m{12.5cm}|}
\hline
Heuristics & Detailed Descriptions\\ \hline
$T_1(d_0, n, lb, ub)$ & \scriptsize The number of circles $m$ is first chosen uniformly randomly from the set $[n]$. The radius and center of the circles are given by $\{r_i \sim U(lb, ub): i\in [m]\}$ and $\{c_i \sim U(r_i, d_0-r_i) \times U(r_i, d_0-r_i): i \in [m]\}$, respectively. \\ \hline
$T_2(d_0, n, t)$ & \scriptsize The number of cracks $m$ is first chosen uniformly randomly from $[n]$. The starting and end locations (2D) are $\{s_i \sim U(0,d_0) \times U(0,d_0): i \in [m]\}$ and $\{e_i \sim U(0,d_0) \times U(0,d_0): i \in [m]\}$, respectively. The thickness of the line $\{t_i, i \in [m]\}$ is chosen uniformly from the set $[t]$. The cracks are then generated using \texttt{cv2.line()} function with the above parameters.\\ \hline
$F(\bv; h, h_1, h_2)$ & \scriptsize Define $\bM_1 \in \reals^{h_1 \times h}$ and $\bM_2 \in \reals^{h_2 \times h_1}$ with $M_1^{(ij)} \sim \cN(1, 1)$ and $M_2^{(ij)} \sim \cN(0, 1)$. Then, the process sensing signal is generated by $F(\bv) = \text{logit}(\lvert \bM_2 \ \text{ReLU}\parent{\bM_1\bv}\rvert) \in \reals^{h_2}$.\\ \hline
\end{tabular}
\label{apd: tab-heuristics}
\end{table}

\subsection{Offline Training \& Baseline Methods} \label{apd: simulation-arch}

In the offline training, the feature extractor model used for simulated quality image is the same as the encoder model used in the data generation process. To extract features from process signature signals which have dimension $1 \times 16$, we utilize an 1D convolutional neural network. It consists of two convolutional layers (\texttt{Conv1d}) with 128 and 64 filters, respectively, each followed by ReLU activation and 1D max pooling (\texttt{MaxPool1d}). The output is passed through three fully connected layers that downsample the features to 64, 32 and then the configurable hidden diemsions, respectively, with ReLU activations applied to the only first two fully connected layers.

To train the feature extractor models for simulated process signature signals and quality images. We use the RMSProp \citep{graves2013generating} optimizer with learning rate 0.1 for 500 epochs in a full batch manner. Note that the window size to estimate the canonical correlation coefficient is set to $n=25$ as introduced in Section 4. The training and evaluation procedure for baseline methods are discussed in the following:
\begin{itemize}
    \item \textbf{Classification-based Model} (\texttt{CLS}): For the classification-based model, we randomly select 250 data points from the simulated abnormal process for training, representing approximately $5\%$ of the size of the normal data. This imbalanced setting reflects practical scenarios where abnormal samples are limited. The classification neural network uses the same feature extractor architecture for process signature signals but employs a classification head composed of three fully connected layers with output dimensions of 16, 16, and 1, respectively. The final layer applies a sigmoid activation function to produce a probability score. The model is trained using RMSProp \citep{graves2013generating}, with the Random Over-Sampling Examples (ROSE) technique \citep{RJ-2014-008} employed to address the class imbalance between normal and abnormal samples. The learning rate is set to $0.001$ and the model is trained with batch size of 256 for 500 epochs.

To evaluate performance, each individual process signature signal within a window of size 25 is evaluated independently by the model. A window is classified as abnormal if the number of detected abnormal signals exceeds a threshold $T \in [25]$. Since no validation set is available, we evaluate the model across all possible thresholds and report the average performance. Note that this classification-based approach is not exactly fair compared to the proposed method, as the latter does not necessarily require abnormal samples for training.
    \item \textbf{PCA + $T^2$:} We extract features from raw process signature signals by fitting a PCA model from the normal process training dataset, selecting the number of principal components such that at least $90\%$ of the total variance is captured. A $T^2$ control chart of those features is then constructed for online monitoring of each window with a size of $n = 25$.
    \item \textbf{PLS + $T^2$:} To develop a Partial Least Squares (PLS) model for feature extraction, we first flatten the simulated quality image as the response variable on the training set. A regression model is then fitted to learn the mapping from the process signature signals to this response in the training set. During evaluation, features from the test set are extracted using the trained model and assessed using a $T^2$ control chart.
\end{itemize}
It is important to note that for both PCA-based and PLS-based methods, the input data are standardized prior to model fitting. This ensures that each feature contributes equally to the model by removing the influence of scale differences, thereby improving the reliability and accuracy of the learned representations.

%% file: src/appendix/case.tex
\section{Case Study}\label{apd: case-study}

\subsection{Laser Material Deposition Experiments}
\label{apd: case-data-collection}
\begin{figure}[t]
    \centering
\includegraphics[width=1.0\linewidth]{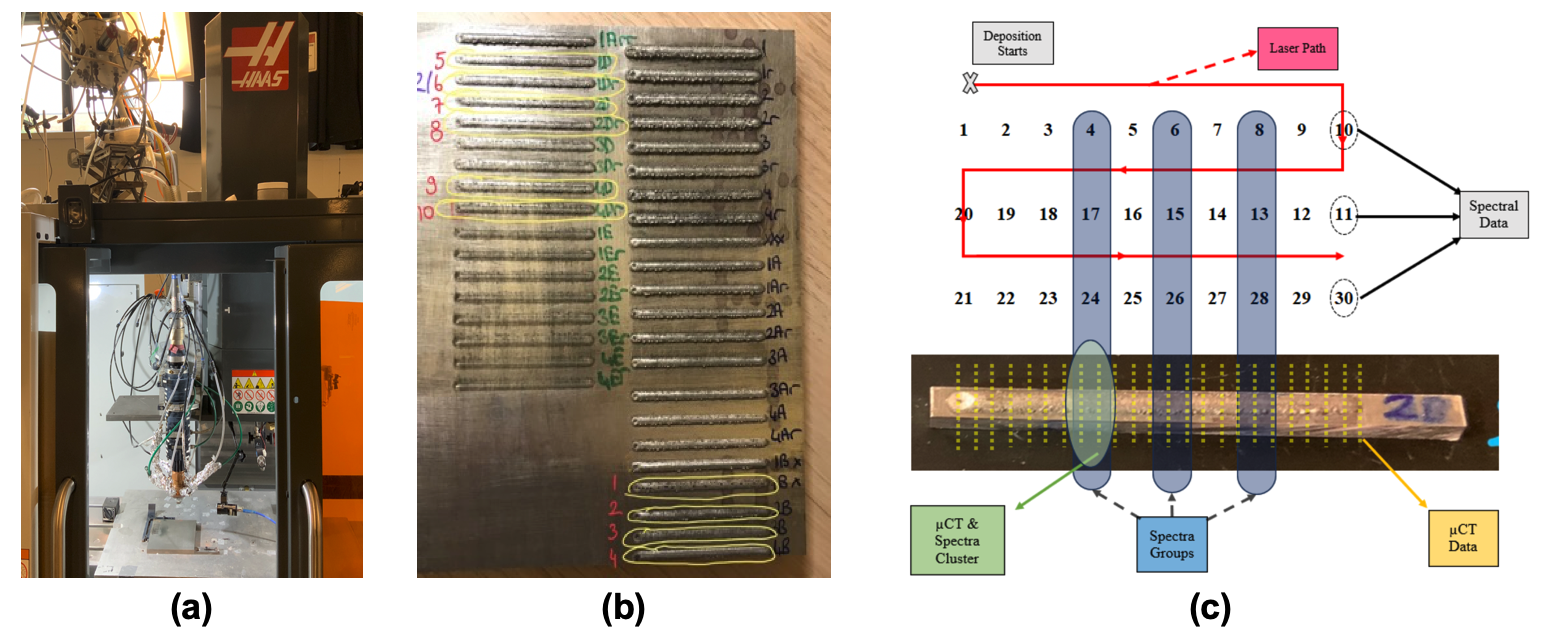}
    \caption{(Figure 7 revisited) An overview of the DMD case study experimental setup is shown in (a), along with example finished parts in (b), and an illustration of the procedures for collecting in-situ spectra and offline CT scans in (c). In short, the printed part is a single line as shown in (b), and the laser path is represented by the red line in (c). \label{apd: fig-example-alignment}}
\end{figure}

Laser material deposition (LMD), also referred to as direct metal deposition (DMD), offers several advantages over other industrialized metal additive manufacturing methods, such as laser powder bed fusion and binder jetting. These advantages include higher build rates, material and build envelope flexibility, and easier process control.
LMD systems consist of several key components: a high-power laser (exceeding 1000 W), a gantry system (CNC or robotic arm), a powder or wire feeder, gas pumps (for shielding and material delivery), and an optical column. Due to its complexity, DMD systems have a large physical footprint. This footprint can be enhanced with external sensors such as CCD cameras, pyrometers, and spectrometers, which enable improved monitoring and control. Figure \ref{apd: fig-example-alignment} (a) provides an illustration of the DMD system setup.

Experiments are carried out with the Aluminum 7075 alloy. Powders were supplied from various vendors and various morphologies, and mechanically mixed. While spherical powders are ideal for additive manufacturing, non-ideal morphologies were intentionally used in this study to induce process defects for tracking during final analysis. The baseplate material was also Aluminum 7075 to eliminate variability from introducing a different substrate material. Its thickness, approximately 1 cm (3/8 inch), served as a heat sink to prevent buckling during laser activity. To ensure effective heat transfer, the baseplate surface was cleaned with a metal brush to remove the oxidation layer, which occurs naturally in the Al surfaces.

The DMD setup was retrofitted from a HAAS Vf2 machine with the required gas, optics, and cooling sections. Detailed parameters are presented for the reader's convenience:
\begin{itemize}
    \item \textbf{Laser:}
    \begin{itemize}
        \item Trumpf HLD 4002
        \item 4000 W maximum power
        \item Yb:YAG -- 1030 nm
        \item BPP 8 mm$\cdot$mrad
        \item Effective laser diameter range: 0.6--1.5 mm (0.02--0.06 inches)
    \end{itemize}

    \item \textbf{Powder Deposition:}
    \begin{itemize}
        \item 3 different powder canisters
        \item Adjustable DC motor shaft driven
        \item Usage of delivery gas to enhance powder flow
    \end{itemize}

    \item \textbf{Gas:}
    \begin{itemize}
        \item Adjustable shielding gas
        \item Adjustable delivery gas
        \item Adjustable purging gas to protect optics
    \end{itemize}

    \item \textbf{CNC Machine:}
    \begin{itemize}
        \item Build Volume:
        \begin{itemize}
            \item Width: 30 cm ($\approx$ 12 inches)
            \item Depth: 20 cm ($\approx$ 8 inches)
            \item Height: 10 cm ($\approx$ 4 inches)
        \end{itemize}
        \item Deposition speed range: 1--5 mm/s (0.05--0.2 ips)
    \end{itemize}

    \item \textbf{Attachable External Sensors:}
    \begin{itemize}
        \item High-speed CCD camera
        \item Optical emission spectra collector
        \item Pyrometer
        \item Thermocouple
    \end{itemize}

\end{itemize}
In addition to the above devices, National Instruments USB6343 and Windows Desktop are used for the production control environment. The detailed parameter configurations for the two processes that we experiment on are provided in Table \ref{apd: tab-real-data-description}.
\begin{table}[t]
\setlength{\aboverulesep}{0pt}
\setlength{\belowrulesep}{0pt}
  \centering
  % \small
    \caption{(Table 2 revisited) Process parameters and description on collected spectra and CT scan dataset. \label{apd: tab-real-data-description}}
  \scalebox{0.63}{
    \begin{tabular}{c|ccc|cc|cc|cc}
    \toprule
    \multirow{2}[6]{*}{\tb{Processes}} &\multicolumn{3}{c}{\tb{Process Configuration}} & \multicolumn{2}{c}{\tb{Data Dimension}} & \multicolumn{2}{c}{\tb{Data Dimension (processed)}} & \multicolumn{2}{c}{\tb{Aligned Dataset}}\\\cmidrule{2-10}
    & \makecell{Power \vspace{-0.2cm}\\(W)} & \makecell{Speed \vspace{-0.2cm}\\(mm/s)} & \makecell{Diameter \vspace{-0.2cm}\\(mm)} & \makecell{Spectra\vspace{-0.2cm}\\($L \times \lambda$)} & \makecell{CT scans\vspace{-0.2cm}\\($L \times C \times H \times W$)} & \makecell{Spectra\vspace{-0.2cm}\\($L \times \lambda$)} & \makecell{CT scans\vspace{-0.2cm}\\($L \times C \times H \times W$)} & \makecell{Training\vspace{-0.2cm}\\size} & \makecell{Testing\vspace{-0.2cm}\\size}\\\hline
    Normal & 1750 & 21 & 1 & $6759 \times 2038$ & $4751 \times 1 \times 674 \times 1000$ & $681 \times 32$ & $4751 \times 1 \times 250 \times 730$ & 89 & 40\\
    Abnormal & 1800 & 15 & 1 & $7376 \times 2038$ & $4046 \times 1 \times 674 \times 1000$ & $942 \times 32$ & $4046 \times 1 \times 250 \times 730$ & 135 & 40\\
    \bottomrule
    \end{tabular}
    }
\end{table}

\subsection{Data Collection \& Processing} \label{apd: case-data}

\subsubsection{Spectra Process Signals}
The spectrometer used in this project is equipped with a 10-µm entrance aperture, a holographic UV/VIS grating with a groove density of 1200 lines per millimeter, and a 2048-element CCD array detector. It provides a spectral resolution of 0.05 nm and records data at 10-ms intervals. Spectra acquisition is initiated immediately before deposition and terminated manually at the end of the experiment. The start and end times are later verified by inspecting the signal intensity, with irrelevant entries removed prior to analysis. In this setup, the collimator lens is positioned 5 mm above the baseplate to capture light–material interactions at the location where the signal is expected to be strongest. As the collimator is rigidly mounted to the DMD head, spatial variance in the collected spectral signals is minimized. This data collection procedure produces a spectra dataset of dimension $6759 \times 2038$ for the normal process, corresponding to 6,759 sampling locations across 2,038 wavelengths. For the abnormal process, the dimension of the data is $7376 \times 2038$, where the discrepancy is due to different process configurations.

To process the raw spectra signals, we only keep the first scan shown in Figure \ref{apd: tab-real-data-description} (c) in the laser path. This choice is motivated by the fact that the first scan is least affected by residual heating from preceding scans. Since we focus exclusively on the first scan, the spatial dimension is truncated to correspond to its start and end. These points are determined by tracking active signals, defined as average intensities across all possible wavelengths exceeding a specified threshold. The thresholds are set to 1000 a.u. for the normal process and 1020 a.u. for the abnormal process, respectively. In addition, the material under study is AL7075, whose relevant wavelength range is 394–398 nm. Accordingly, we clip the spectra to 393.829-398.168 nm, which significantly reduces unnecessary noise. This yields a wavelength dimension of 96, which we further downsample to 32 by selecting equally spaced points.
\begin{figure}[t]
    \centering
    \includegraphics[width=1.0\linewidth]{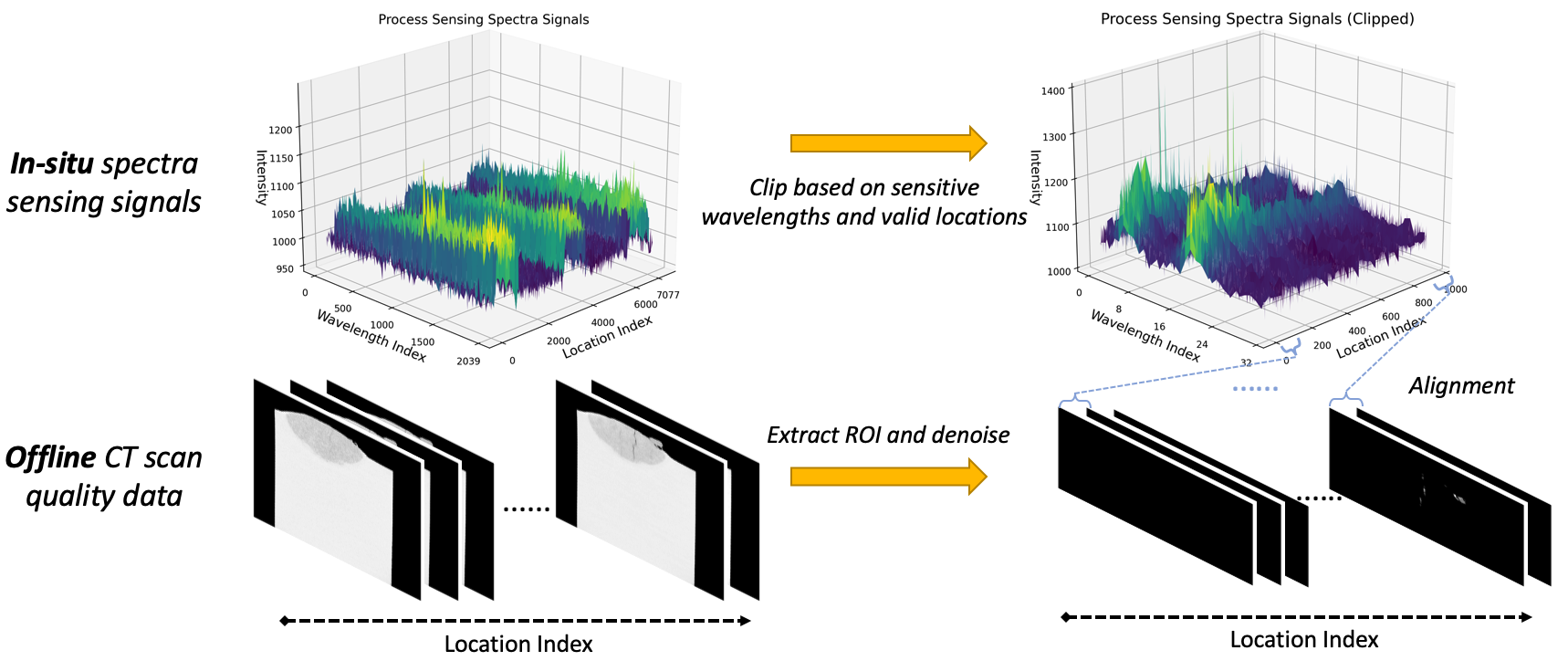}
    \caption{Illustration of the data processing procedure: the spectra signals are clipped according to relevant spatial locations and sensitive wavelength ranges, while the quality CT scans are first cropped to the regions of interest and then denoised to a binary format.}
    \label{apd: fig-more-examples}
\end{figure}

\subsubsection{CT Scans Quality Images}
Upon completion of the deposition experiments, we started post-processing. First, loose powders on the deposition surface were cleaned with an air gun and a soft brush. The Figure \ref{apd: tab-real-data-description} (b) was captured at this step. After discussion with MicroCT experts, it was noted that the baseplate thickness was too much for adequate X-ray penetration. To address this, the backside of the deposition was machined down to approximately 5 mm. Excessive thinning was avoided to preserve the under-melted area, which is critical for analyzing the bonding between the first deposition layer and the baseplate. Individual stripes, which are produced under different process configurations, were scanned using a MicroCT machine, capturing 2D slices at intervals of approximately 10 microns. The MicroCT data were compiled into a video folder containing sequential 2D frames along the deposition. This yields a sequence of CT scans of dimension $4751 \times 1 \times 674 \times 1000$ for normal process, representing 4751 gray-scale images with dimension $674 \times 1000$. Similarly, the abnormal process data has dimensions $4046 \times 1 \times 674 \times 1000$.

The original images contain excessive white background, which is irrelevant for defect identification and introduces unnecessary noise. We retain only the top $250 \times 730$ portion, as most defects appear in this region in our experiment. In the original image, this corresponds to the region with $0:250$ and $120:850$ in the height and width dimensions, respectively. In addition, the images are denoised by binarizing with a threshold automatically selected by Otsu's method \citep{otsu1975threshold} and inverting the pixel values so that defect pixels have a value of 1. Figure \ref{apd: fig-more-examples} provides an overview of this data processing procedure.

\subsubsection{Data Alignment}
After processing the spectra and CT scans according to the aforementioned procedures, the data are aligned based on their corresponding spatial locations. In this case study, the produced part has a length of 35 mm. To align these data, a step size of 0.27 mm and 0.20 mm for the normal and abnormal processes, respectively. These step sizes compensate for the differences in data collection frequencies between the two configurations and ensure that each aligned tuple contains 5 spectra and 22 CT scan images for both processes.

\subsection{Model Architecture \& Offline Training}\label{apd: case-arch}
We implemented a convolutional neural network \citep{lecun2015deep} architecture to process the input quality CT scans. The feature extraction network begins with 2 layers of convolutional modules (\texttt{Conv2d}) with kernel size $7 \times 7$ and padding $3 \times 3$, followed by a $3 \times 3$ maxpooling layer (\texttt{maxpool2d}). The output channel dimensions for the two layers are 16 and 8, respectively. The fully connected portion of the network consists of four linear layers, with output sizes 2560, 128, 64, and the hidden feature dimension, respectively.

The feature extractor for spectra process signature signals follows a similar architecture introduced by \citet{sun2022situ}. Specifically, we first designed a one-dimensional convolutional network to extract temporal features from the spectral signals. The first convolutional layer (\texttt{Conv1d}) applies 256 filters with a kernel size of 6, followed by a second convolutional layer with 32 filters of the same kernel size. Two maxpooling layers (\texttt{maxpool2d}) with a pooling size of 3. Then, we employed a two-layer long short-term memory (LSTM) network. The first LSTM layer receives 32-dimensional inputs and projects them into a 128-dimensional hidden state, while the second LSTM layer further processes the sequence and maps the 128-dimensional hidden state into the desired feature dimension.

The overall DCCA-based feature extractors are designed to have a hidden dimension of 6, and are optimized using RMSProp with a learning rate $0.00005$ for 500 epochs. Note that this smaller learning rate is particularly useful for stably training a DCCA model with small window size \citep{andrew2013deep}. The window size is set to $n = 8$. To validate its training performance and also select the online decision threshold, a validation set that is $20\%$ of the original data is held unseen during training.